\documentclass[twocolumn,secnumarabic,amssymb,nobibnotes,
superscriptaddress,nofootinbib,aps,prd]{revtex4-2}

\usepackage{natbib}
\usepackage{xcolor}
\usepackage{adjustbox}
\usepackage{amsmath}
\usepackage{graphicx}
\usepackage{subcaption}
\usepackage{booktabs}
\usepackage{amsfonts}
\usepackage{nicefrac}
\usepackage{microtype}
\usepackage{multirow}
\usepackage{makecell}
\usepackage{array}
\usepackage[hidelinks]{hyperref}
\usepackage{orcidlink}
\usepackage{float}
\usepackage{comment}
\usepackage[capitalise]{cleveref}
\usepackage{caption}
\begin{document}

\title{Reaction-constrained composition \(g\)-modes in neutron stars with antikaon condensates, hyperons, and \(\Delta(1232)\) resonances}

\author{Prashant Thakur \orcidlink{0000-0003-4189-6176}}
\email{prashant@yonsei.ac.kr}
\affiliation{Department of Physics, Yonsei University, Seoul, 03722, South Korea}

\author{Ishfaq Ahmad Rather~\orcidlink{0000-0001-5930-7179}}
\email{rather@astro.uni-frankfurt.de}
\affiliation{Institut f\"{u}r Theoretische Physik, Goethe Universit\"{a}t,
Max-von-Laue-Str.~1, D-60438 Frankfurt am Main, Germany}

\begin{abstract}
We investigate core composition \(g_1\) modes of cold, nonrotating
neutron stars containing antikaon condensates, hyperons, and
\(\Delta(1232)\) baryons. To our knowledge, we present the first
full-general-relativistic calculation of the continuous-composition
core \(g_1\)-mode frequency and gravitational-wave damping time for
neutron stars containing a \(K^-\) condensate. Using the BigApple relativistic mean-field equation of state, we calculate mode
frequencies, gravitational-wave damping times, and
frozen-composition tidal overlaps in full general relativity. We also
introduce a species-resolved Ledoux decomposition, independently
validated through mode-frequency sensitivities. To account for
composition-changing reactions, we compare fully frozen matter with a
fast-\(K\) limit for \(n\leftrightarrow p+K^-\) and a
strong-equilibrium limit for the \(\Delta\) quartet. Fast-\(K\)
equilibration retains approximately \(36\%\)--\(44\%\) of the peak
local kaon buoyancy. At the terminal configurations, the mode
frequencies retain \(65.7\%\)--\(73.4\%\) of their
frozen-composition values, while the gravitational-wave damping times
increase by factors of \(14.4\)--\(31.8\). The kaon-associated mode
nevertheless remains above the nucleonic band. By contrast, strong
\(\Delta\) equilibration removes most of the direct
\(\Delta\)-induced enhancement, returning the \(N\Delta\) mode toward
the nucleonic band, while the high-frequency \(NY\Delta\) branch
survives through the frozen \(\Lambda\) gradient. Eigenfunction node
counts and overlaps confirm that these changes occur along the same
continuous \(g_1\) branch. Representative DD-ME2 calculations
reproduce the same reaction-channel hierarchy. The resulting direct
full-GR, frozen-composition tidal phase shifts satisfy
\(|\Delta\Phi_{g_1}|\leq1.410\times10^{-3}\) rad, approximately a
factor of \(21\) below the \(0.03\)-rad favorable-event sensitivity
scale quoted for the Einstein Telescope. These results show that an
exotic species produces a distinct composition \(g\) mode only when
its associated composition gradient, or a coupled slowly
equilibrating gradient, survives over the oscillation period.

\end{abstract}

\maketitle

\section{Introduction}

Neutron stars provide a unique laboratory for cold, strongly interacting matter at densities several times nuclear saturation, beyond the reach of terrestrial experiments. Their macroscopic properties—including masses, radii, moments of inertia, and tidal deformabilities—are governed by the dense-matter equation of state (EOS) \cite{Lattimer:2000nx,Lattimer:2015nhk,Oertel:2016bki}. Multimessenger observations, particularly the tidal constraints from GW170817 \cite{LIGOScientific:2017vwq} and the NICER mass–radius measurements of PSR J0030+0451 and PSR J0740+6620 \cite{Riley:2021pdl,Miller:2021qha}, have substantially restricted the allowed EOS. The microscopic composition of the inner core, however, remains uncertain. Exotic degrees of freedom such as hyperons, $\Delta(1232)$ isobars, antikaon condensates, or deconfined quarks may emerge at supranuclear densities. Because these components can produce comparable modifications of the EOS, distinct core compositions may yield nearly indistinguishable static observables, leaving the internal particle content underdetermined by mass, radius, and tidal-deformability measurements alone.

Neutron-star asteroseismology provides such a probe through the characteristic frequencies and damping times of stellar oscillations \cite{Andersson:1997rn,Kokkotas:1999bd,Thakur:2025qwl}. The nonradial spectrum includes the fundamental $f$ mode, pressure $p$ modes, gravity $g$ modes, rotational $r$ modes, and spacetime $w$ modes \cite{thorne1967,Lindblom:1983ps,Detweiler:1985zz,Kokkotas:1999bd}. These mode families probe different aspects of stellar structure, including compressibility, composition, rotation, and spacetime curvature. Since several of them couple to gravitational radiation, their frequencies and damping times can constrain the EOS and internal composition beyond equilibrium observables \cite{Andersson:1997rn,Benhar:2004xg,Doneva:2015jba}. The enhanced sensitivity of third-generation gravitational-wave observatories, particularly the Einstein Telescope and Cosmic Explorer, is expected to substantially improve the prospects for detecting neutron-star oscillation signatures and developing gravitational-wave asteroseismology \cite{Kokkotas:2000up,Andersson:2021qdq}.

Among the different oscillation families, composition $g$-modes are particularly promising probes of the microscopic content of neutron-star cores. Their restoring force is buoyancy, which arises when a fluid element is displaced on a timescale shorter than that required for weak interactions to restore local $\beta$ equilibrium. The displaced element therefore retains, to a good approximation, its original particle fractions and develops a density contrast relative to the surrounding equilibrium matter. The strength of the buoyancy is determined by the difference between the adiabatic sound speed evaluated at frozen composition and the equilibrium sound speed along the $\beta$-equilibrated stellar background \cite{Lai:1993di,Tran:2022dva}. Because this difference depends directly on the local composition gradients, the onset of a new particle species or phase can produce pronounced features in the buoyancy profile and corresponding changes in the $g$-mode spectrum. Such behavior has been reported for the appearance of deconfined quarks~\cite{Zhao:2022toc}, hyperons~\cite{Tran:2022dva}, and $\Delta(1232)$ baryons~\cite{Sun:2026agt}. Composition $g$-modes can therefore distinguish stellar models that remain nearly degenerate in their masses, radii, and tidal deformabilities.

The $g$-mode spectrum is also sensitive to the thermodynamic and reaction state of the star. Finite temperature and composition-changing weak reactions can suppress or reshape the buoyancy, particularly in proto-neutron stars and binary-merger remnants~\cite{Counsell:2023pqp,Zhao:2025pgx}. In inspiralling binary neutron-star systems, these modes may become observable through their resonant interaction with the dynamical tide~\cite{1994ApJ...426..688R,Lai:1993di,Andersson:2019dwg}. As the orbital frequency increases, the tidal forcing can sweep through a $g$-mode resonance, transferring energy and angular momentum to the star and producing a phase correction in the gravitational waveform. Although the coupling is generally weak for slowly rotating stars in quasicircular binaries, it can be enhanced by stellar rotation, orbital eccentricity, and a sufficiently large tidal-overlap integral. The improved sensitivity of third-generation gravitational-wave observatories, particularly the Einstein Telescope and Cosmic Explorer, may therefore bring these resonant signatures within observational reach~\cite{ET:2019dnz,Branchesi:2023mws,Evans:2021gyd}.

Despite this progress, two unresolved issues motivate the present
work. First, although antikaon (\(K^-\)) condensation has been studied
extensively in neutron-star matter, its effect on composition
\(g\)-modes has not, to our knowledge, been calculated. Existing
composition-mode studies have considered quarks, hyperons, and
\(\Delta\) isobars \cite{Zhao:2022toc,Tran:2022dva,Sun:2026agt}, but
not a Bose condensate of antikaons. Second, the conventional assumption
that all particle fractions remain frozen during an oscillation is a
reaction-timescale limit rather than a universal property. The
\(\Delta\) fractions can be rearranged through strong reactions such
as \(nn\leftrightarrow p\Delta^-\), whereas kaon conversion can proceed
through the nonleptonic weak reaction
\(n\leftrightarrow p+K^-\) \cite{Chatterjee:2007qs}. The free-space
\(\Delta\) width, \(\Gamma_\Delta\simeq117\) MeV
\cite{ParticleDataGroup:2024cfk}, indicates the underlying strong
interaction scale but does not by itself determine the relaxation rate
in cold, degenerate stellar matter. The survival of the corresponding
buoyancy must therefore be tested using appropriate limiting
composition constraints.

We address these questions within a single full-general-relativistic
framework. Using the Lindblom--Detweiler formulation
\cite{Lindblom:1983ps,Detweiler:1985zz}, we calculate the core
\(g_1\)-mode frequencies and gravitational-wave damping times of
BigApple matter containing hyperons, \(\Delta\) baryons, and an
antikaon condensate. We introduce a species-resolved Ledoux
decomposition
\cite{1947ApJ...105..305L,1983ApJ...268..837M,1992ApJ...395..240R}
to identify the microscopic origin of the buoyancy and compare the
fully frozen, fast-\(K\), and strong-equilibrium \(\Delta\) limits.
The comparison shows that the kaon-associated mode survives in both
kaon-reaction limits, whereas the direct \(\Delta\)-induced enhancement
is strongly suppressed when the \(\Delta\) quartet maintains strong
equilibrium. We additionally verify the mode identity and
species attribution independently and estimate the frozen-limit tidal
phase shifts relative to the favorable-event sensitivity scale of the
Einstein Telescope \cite{Gittins:2026ntx}.

The paper is organized as follows. In Sec.~\ref{sec:eos}, we introduce
the equation of state and its extensions containing hyperons,
\(\Delta\) baryons, and antikaon condensation, together with the
full-general-relativistic oscillation formalism and species-resolved
Ledoux decomposition. In Sec.~\ref{res}, we present the buoyancy
profiles, mass--radius sequences, \(g_1\)-mode frequencies and damping
times, and the contributions of individual particle species. We then
compare the fully frozen, fast-\(K\), and strong-equilibrium
\(\Delta\) limits, validate the mode identity and species attribution,
and examine resonant tidal excitation during binary inspiral.
Section~\ref{sec:conclusions} summarizes our conclusions.

\section{Methodology}\label{sec:eos}

\subsection{Equation of state}

We describe uniform stellar matter with the nonlinear relativistic
mean-field (RMF) model in the BigApple parametrization
\cite{Fattoyev:2020cws}. The nucleonic Lagrangian density is
\begin{equation}
\mathcal{L}_{\rm BA}^{(N)}
 =\mathcal{L}_{N}
 +\mathcal{L}_{\sigma\omega\rho}
 +\mathcal{L}_{\ell},
\label{eq:L_total}
\end{equation}
where the baryon and lepton pieces and the nucleon effective mass are
\begin{equation}
\begin{aligned}
\mathcal{L}_{N} ={}&\sum_{N=n,p}\bar{\psi}_{N}
 \Big[\gamma_{\mu}\!\left(
 i\partial^{\mu}-g_{\omega N}\omega^{\mu}
 -g_{\rho N}I_{3N}\rho_{3}^{\mu}\right)\\
 &\qquad\quad -\,m_N^{*}\Big]\psi_{N},\\[3pt]
\mathcal{L}_{\ell} ={}&\sum_{\ell=e,\mu}
 \bar{\psi}_{\ell}
 \left(i\gamma_{\mu}\partial^{\mu}-m_{\ell}\right)\psi_{\ell},\\[3pt]
m_N^{*} ={}& m_N-g_{\sigma N}\sigma ,
\end{aligned}
\label{eq:baryon_lepton}
\end{equation}
and the complete nonlinear meson sector is
\begin{equation}
\begin{aligned}
\mathcal{L}_{\sigma\omega\rho}={}&
 \frac{1}{2}\partial_{\mu}\sigma\,\partial^{\mu}\sigma
 -\frac{1}{2}m_{\sigma}^{2}\sigma^{2}
 -\frac{\kappa}{3!}(g_{\sigma N}\sigma)^{3}
 -\frac{\lambda}{4!}(g_{\sigma N}\sigma)^{4}\\
&-\frac{1}{4}\Omega_{\mu\nu}\Omega^{\mu\nu}
 +\frac{1}{2}m_{\omega}^{2}\omega_{\mu}\omega^{\mu}
 +\frac{\zeta}{4!}g_{\omega N}^{4}
  (\omega_{\mu}\omega^{\mu})^{2}\\
&-\frac{1}{4}\boldsymbol{R}_{\mu\nu}\!\cdot\!
  \boldsymbol{R}^{\mu\nu}
 +\frac{1}{2}m_{\rho}^{2}
  \boldsymbol{\rho}_{\mu}\!\cdot\!\boldsymbol{\rho}^{\mu}\\
&+\Lambda_v\, g_{\omega N}^{2}g_{\rho N}^{2}
 (\omega_{\mu}\omega^{\mu})
 (\boldsymbol{\rho}_{\nu}\!\cdot\!\boldsymbol{\rho}^{\nu}) .
\end{aligned}
\label{eq:meson}
\end{equation}

Here \(\Omega_{\mu\nu}\) and \(\boldsymbol{R}_{\mu\nu}\) are the
\(\omega\)- and \(\rho\)-meson field tensors. In uniform matter only
the temporal mean fields \(\sigma\), \(\omega_0\), and \(\rho_{03}\)
remain nonzero. The canonical BigApple parameters and the saturation properties they reproduce are collected in \cref{tab:bigapple}. We use the isovector vertex \(g_{\rho N}I_{3N}\), with \(I_{3p}=+1/2\),
\(I_{3n}=-1/2\) and \(g_{\rho N}=\sqrt{200.5562}\); no additional factor of \(1/2\) is introduced.

\begin{table}[t]
\centering
\caption{BigApple model parameters (upper block) and the saturation
properties they reproduce (lower block). Squared meson--nucleon
couplings are dimensionless; \(\kappa\) has units of MeV and
\(\lambda,\zeta,\Lambda_v\) are dimensionless.}
\label{tab:bigapple}
\begin{ruledtabular}
\begin{tabular}{l r l}
Quantity & Value & Unit\\
\hline
$m_N$              & $939$        & MeV\\
$m_\sigma$         & $492.730$    & MeV\\
$m_\omega$         & $782.500$    & MeV\\
$m_\rho$           & $763.000$    & MeV\\
$g_{\sigma N}^{2}$ & $93.5074$    & --\\
$g_{\omega N}^{2}$ & $151.6839$   & --\\
$g_{\rho N}^{2}$   & $200.5562$   & --\\
$\kappa$           & $5.20326$    & MeV\\
$\lambda$          & $-0.021739$  & --\\
$\zeta$            & $0.000700$   & --\\
$\Lambda_v$        & $0.047471$   & --\\
\hline
$n_0$              & $0.155$      & fm$^{-3}$\\
$E/A$              & $-16.344$    & MeV\\
$K$                & $227.001$    & MeV\\
$J$                & $31.315$     & MeV\\
$L$                & $39.800$     & MeV\\
\end{tabular}
\end{ruledtabular}
\end{table}

For the antikaon-condensed sequences \cite{Chatterjee:2007qs}, a single \(s\)-wave \(K^-\)
field is added through
\begin{equation}
\begin{aligned}
\mathcal{L}_{K^-}={}&
 \left(D_{\mu}^{(-)}K^-\right)^{\dagger}
 D^{(-)\mu}K^-
 -m_K^{*\,2}K^{-\dagger}K^-,\\[3pt]
m_K^{*}={}&m_K-g_{\sigma K}\sigma ,
\end{aligned}
\label{eq:LK}
\end{equation}
with the covariant derivative
\begin{equation}
D_{\mu}^{(-)}K^-
 =\left(
 \partial_{\mu}
 -i g_{\omega K}\omega_{\mu}
 -\tfrac{i}{2}g_{\rho K}\rho_{\mu}^{3}
 \right)K^- .
\label{eq:Dmu}
\end{equation}
For \(K^-\propto e^{-i\omega_{K^-}t}\), this convention gives the in-medium dispersion relation
\begin{equation}
\omega_{K^-}
 =m_K^{*}-g_{\omega K}\omega_0
 -\tfrac{1}{2}g_{\rho K}\rho_{03}.
\label{eq:omegaK}
\end{equation}
We take \(m_K=497.5\) MeV,
\(g_{\omega K}=g_{\omega N}/3\), and
\(g_{\rho K}=g_{\rho N}\) (\cref{tab:exotic}). The scalar coupling is calibrated at saturation from
\begin{equation}
U_K(n_0)
 =-g_{\sigma K}\sigma_0-g_{\omega K}\omega_0,
\label{eq:UK}
\end{equation}
using \(U_K=-100,-120,-140\), and \(-160\) MeV; the \(U_K=0\) sequence is the condensate-free \(npe\mu\) control rather than an additional kaon calculation. Condensation begins when
\(\mu_e=\mu_n-\mu_p=\omega_{K^-}\). For the zero-momentum condensate, the direct thermodynamic contribution is
\begin{equation}
\varepsilon_K=m_K^{*}\,n_K,\qquad P_K=0,
\label{eq:epsK}
\end{equation}
while its effect on the total pressure enters through the self-consistent meson fields.

The hyperonic extension contains the six strange members of the baryon octet,
\(Y=\{\Lambda,\Sigma^-,\Sigma^0,\Sigma^+,\Xi^-,\Xi^0\}\), and is described by
\begin{equation}
\begin{aligned}
\mathcal{L}_{Y}={}&
\sum_Y\bar{\psi}_{Y}
\Big[\gamma_{\mu}\!\left(
i\partial^{\mu}-g_{\omega Y}\omega^{\mu}
-g_{\rho Y}I_{3Y}\rho_{3}^{\mu}\right.\\
&\left.\qquad\quad
-g_{\phi Y}\phi^{\mu}
\right)-m_Y^{*}\Big]\psi_Y
+\mathcal{L}_{\phi},\\[3pt]
m_Y^{*}={}&m_Y-g_{\sigma Y}\sigma ,
\end{aligned}
\label{eq:LY}
\end{equation}
where the hidden-strangeness \(\phi\) sector is
\begin{equation}
\mathcal{L}_{\phi}
 =-\frac{1}{4}\Phi_{\mu\nu}\Phi^{\mu\nu}
 +\frac{1}{2}m_{\phi}^{2}\phi_{\mu}\phi^{\mu},
 \qquad m_\phi=1019.461~{\rm MeV}.
\label{eq:Lphi}
\end{equation}
Here \(\Phi_{\mu\nu}=\partial_\mu\phi_\nu-\partial_\nu\phi_\mu\). The SU(6) vector-coupling relations are \cite{Schaffner:1996ge}
\begin{equation}
\begin{gathered}
g_{\omega\Lambda}=g_{\omega\Sigma}
 =\tfrac{2}{3}g_{\omega N},\qquad
g_{\omega\Xi}=\tfrac{1}{3}g_{\omega N},\\
g_{\rho\Lambda}=0,\qquad
g_{\rho\Sigma}=g_{\rho\Xi}=g_{\rho N},\\
g_{\phi\Lambda}=g_{\phi\Sigma}
 =-\tfrac{\sqrt{2}}{3}g_{\omega N},\qquad
g_{\phi\Xi}=-\tfrac{2\sqrt{2}}{3}g_{\omega N},
\end{gathered}
\label{eq:su6}
\end{equation}
with no \(\sigma^*\) meson. The scalar couplings are fixed from the single-particle potentials in symmetric nuclear matter,
\begin{equation}
U_Y^{(N)}(n_0)
 =-g_{\sigma Y}\sigma_0+g_{\omega Y}\omega_0,
\label{eq:UY}
\end{equation}
using the values listed in \cref{tab:exotic}
\cite{Gal:2016boi}.

\begin{table}[t]
\centering
\caption{Adopted antikaon, hyperon, and \(\Delta\) couplings. The
\(K^-\) and hyperon scalar couplings are calibrated from the listed
potentials at saturation density. Hyperon vector couplings follow the
SU(6) relations in Eq.~\eqref{eq:su6}, whereas the \(\Delta\) couplings
are specified by the ratios \(x_{i\Delta}=g_{i\Delta}/g_{iN}\). The
listed \(U_\Delta^{(N)}(n_0)\) is predicted by these ratios.}
\label{tab:exotic}
\begin{ruledtabular}
\begin{tabular}{l l l}
Sector & Quantity & Value\\
\hline
$K^-$
 & $m_K$          & $497.5$~MeV\\
 & $g_{\omega K}$  & $g_{\omega N}/3$\\
 & $g_{\rho K}$    & $g_{\rho N}$\\
 & $U_K(n_0)$      & $-100,-120,-140,-160$~MeV\\
\hline
$\Lambda$ & $U_\Lambda(n_0)$ & $-28$~MeV\\
$\Sigma$  & $U_\Sigma(n_0)$  & $+30$~MeV\\
$\Xi$     & $U_\Xi(n_0)$     & $-14$~MeV\\
\hline
$\Delta$
 & $x_{\sigma\Delta}$ & $1.09$\\
 & $x_{\omega\Delta}$ & $1.05$\\
 & $x_{\rho\Delta}$   & $2.5$\\
 & $g_{\phi\Delta}$   & $0$\\
 & $U_\Delta^{(N)}(n_0)$   & $-90.9$~MeV\\
\end{tabular}
\end{ruledtabular}
\end{table}

For the \(\Delta(1232)\) quartet, the uniform-matter mean-field reduction is 
\begin{equation}
\begin{aligned}
\mathcal{L}_{\Delta}^{\rm MF}={}&
 \sum_{d}
 \bar{\psi}_{d,\alpha}
 \Big[\gamma_{\mu}\!\left(
 i\partial^{\mu}-g_{\omega\Delta}\omega^{\mu}\right.\\
&\left.\qquad
 -g_{\rho\Delta}I_{3d}\rho_{3}^{\mu}
 \right)-m_\Delta^{*}\Big]\psi_d^{\alpha},
\end{aligned}
\label{eq:LDelta}
\end{equation}
where \(d\in\{\Delta^-,\Delta^0,\Delta^+,\Delta^{++}\}\),
\(m_\Delta^{*}=m_\Delta-g_{\sigma\Delta}\sigma\),
\(m_\Delta=1232~{\rm MeV}\), and
\[
I_{3d}=
\left(-\tfrac{3}{2},-\tfrac{1}{2},
+\tfrac{1}{2},+\tfrac{3}{2}\right).
\]
The thermodynamic calculation uses spin degeneracy four. The adopted
ratios \(x_{i\Delta}=g_{i\Delta}/g_{iN}\) are listed in
Table~\ref{tab:exotic}
\cite{Marquez:2022gmu,Cai:2015hya,Ribes:2019kno}. These couplings
predict
\begin{equation}
U_\Delta^{(N)}(n_0)
=-x_{\sigma\Delta}g_{\sigma N}\sigma_0
+x_{\omega\Delta}g_{\omega N}\omega_0
=-90.9~{\rm MeV}.
\label{eq:Udelta}
\end{equation}
Thus, the \(\Delta\) potential is a consequence of the adopted coupling
ratios rather than an independently fitted input. The \(\Delta\) couplings remain uncertain. Within the commonly adopted
window \(x_{\sigma\Delta}\gtrsim1\) and
\(0\leq x_{\sigma\Delta}-x_{\omega\Delta}\leq0.2\)
\cite{Marquez:2022gmu}, increasing \(x_{\sigma\Delta}\) at fixed vector
couplings generally lowers the \(\Delta^-\) onset density and can shift
the departure from the nucleonic branch to lower mass. We do not survey
this coupling dependence here; all \(\Delta\) results below use the
fixed ratios in Table~\ref{tab:exotic}. Imposing strong equilibrium
removes the direct buoyancy associated with frozen \(\Delta\)-fraction
gradients, but the remaining buoyancy and the quantitative global-mode
frequencies can still depend on the adopted couplings.

All active baryons satisfy \(\beta\) equilibrium and charge neutrality,
\begin{equation}
\begin{aligned}
\mu_b={}&\mu_n-q_b\mu_e, \\
\mu_\mu={}&\mu_e,\\
 \sum_b q_b n_b{}&-n_e-n_\mu-n_K=0,
\end{aligned}
\label{eq:beta}
\end{equation}
where \(n_K=0\) for the nonkaonic families and the total baryon density
is \(n_B=\sum_b n_b\). The calculated families are therefore
\(npe\mu\), \(npe\mu+K^-\), \(npe\mu+Y\), \(npe\mu+\Delta\), and
\(npe\mu+Y+\Delta\). At low density, the uniform core EOS is matched to the BPS crust \cite{Baym:1971pw}. If the nucleon effective mass approaches zero before a TOV turning point is reached, the sequence is terminated at the last positive-\(m_N^*\) configuration and identified as an EOS-validity endpoint. The equilibrium and constrained-composition sound speeds used in the oscillation calculation are defined in the species-resolved Ledoux subsection below.

\subsection{Non-radial Oscillations in General Relativity}

To calculate the oscillation spectrum of neutron stars, we employ the
relativistic formulation of linear, non-radial polar perturbations
developed by \citep{thorne1967, Lindblom:1983ps, Detweiler:1985zz}. The equilibrium stellar background is assumed to be static and spherically symmetric, with the metric
\begin{equation}
ds^{2}=-e^{\nu(r)}dt^{2}+e^{\lambda(r)}dr^{2}
+r^{2}(d\theta^{2}+\sin^{2}\theta d\phi^{2}),
\end{equation}
where, $e^{\lambda(r)}=
\left(1-\frac{2Gm(r)}{c^{2}r}\right)^{-1}$.
Small perturbations of the spacetime are described using theeven-parity Regge--Wheeler metric \citep{Regge:1957td,thorne1967},
\begin{equation}
\begin{aligned}
ds^{2} =&
-e^{\nu(r)}
\left[1+r^{l}H_{0}(r)Y_{lm}e^{-i\omega t}\right]dt^{2} \\
&+e^{\lambda(r)}
\left[1-r^{l}H_{0}(r)Y_{lm}e^{-i\omega t}\right]dr^{2} \\
&+r^{2}
\left[1-r^{l}K(r)Y_{lm}e^{-i\omega t}\right]d\Omega^{2} \\
&+2i\omega r^{l+1}H_{1}(r)Y_{lm}e^{-i\omega t}dt\,dr ,
\end{aligned}
\end{equation}
where $H_0$, $H_1$, and $K$ are metric perturbation functions,
$Y_{lm}$ denotes the spherical harmonics, and
$d\Omega^{2}=d\theta^{2}+\sin^{2}\theta d\phi^{2}$.

The fluid displacement associated with the oscillation is represented
by the Lagrangian displacement vector \citep{Lindblom:1983ps,Detweiler:1985zz}
\begin{equation}
\begin{aligned}
\xi^{r}={}&r^{l-1}e^{-\lambda/2}W(r)Y_{lm}e^{-i\omega t}, \\
\xi^{\theta}
={}&-r^{l-2}V(r)\partial_{\theta}Y_{lm}e^{-i\omega t}, \\
\xi^{\phi}
={}&-\frac{r^{l-2}}{\sin^{2}\theta}
V(r)\partial_{\phi}Y_{lm}e^{-i\omega t}.
\end{aligned}
\end{equation}
The Lagrangian pressure perturbation is written as
\[
\Delta p=-r^{l}e^{-\nu/2}X(r)Y_{lm}e^{-i\omega t},
\]
where $X(r)$ is the pressure perturbation variable.

Following the Lindblom--Detweiler formalism
\citep{Lindblom:1983ps,Detweiler:1985zz}, the perturbed Einstein equations
and conservation laws reduce to a set of four coupled first-order
differential equations for the variables $H_1$, $K$, $W$, and $X$.
The remaining perturbation functions $H_0$ and $V$ are obtained
algebraically from these variables. The oscillation equations depend on
the equilibrium stellar structure through the pressure $p$, energy
density $\epsilon$, enclosed mass $m(r)$, and metric functions
$\nu(r)$ and $\lambda(r)$.

For composition $g$ modes, the relevant thermodynamic input is the
difference between the frozen-composition and equilibrium sound speeds.
The frozen-composition sound speed is defined as
\[
c_s^{2}=\left(\frac{\partial p}{\partial \epsilon}\right)_{x_i},
\]
where the derivative is taken at fixed particle fractions $x_i$.
This differs from the equilibrium sound speed,
\[
c_e^{2}=\frac{dp}{d\epsilon},
\]
which is evaluated along the $\beta$-equilibrated stellar sequence.
When $c_s^2=c_e^2$, the matter is locally barotropic and does not
provide a composition-buoyancy restoring force. A nonzero difference
between these two sound speeds produces stable stratification and
supports core $g$ modes
\citep{1994ApJ...426..688R,1992ApJ...395..250G,Zhao:2022toc}.

Regularity of the eigenfunctions is imposed at the stellar center.
The overall normalization of the fluid displacement is fixed by taking $W(0)=1$.
At the stellar surface, the vanishing of the Lagrangian pressure
perturbation gives the outer fluid boundary condition
$X(R)=0$.
For the full general-relativistic calculation, the interior solution is
matched to the exterior spacetime perturbation. Perturbations are taken
to vary as \(e^{-i\omega t}\), with the outgoing Zerilli solution
behaving as \(Z\propto e^{+i\omega r_*}\). A damped quasinormal mode
therefore has \(\operatorname{Im}\omega<0\). The resulting eigenvalue
problem is solved numerically to obtain the complex oscillation
frequencies. The real part of the eigenfrequency gives the oscillation
frequency,
\[
f=\frac{{\rm Re}(\omega)}{2\pi},
\]
and we quote the amplitude e-folding time,
\[
\tau_{g_1}^{\rm GW}=\frac{1}{|\operatorname{Im}\omega|};
\]
the corresponding mode-energy e-folding time is
\(\tau_{g_1}^{\rm GW}/2\).
The interior perturbation equations are integrated on a uniform radial
grid and matched to the exterior Zerilli solution. Representative
resolution-doubling tests change the full-GR quasinormal-mode
frequencies and damping times by at most approximately \(0.6\%\).
After applying the one-sided continuum treatment at composition
thresholds, the species-resolved terms reproduce the total Ledoux term
with a relative closure error below \(10^{-3}\) over the resolved
buoyant region. Thus, unlike the Cowling approximation, the full
general-relativistic treatment provides both the mode frequency and its
damping time due to gravitational-wave emission
\citep{Lindblom:1983ps,Detweiler:1985zz,Andersson:1997rn}.

\subsection{Species-resolved Ledoux buoyancy}

For composition \(g\)-modes, the relevant thermodynamic input is the
difference between the frozen-composition and beta-equilibrium sound
speeds. This buoyancy is the Ledoux buoyancy associated with composition
stratification in neutron-star matter
\citep{1947ApJ...105..305L,1983ApJ...268..837M,1992ApJ...395..240R}. We write the local equation of state as
\begin{equation}
P=P(n_B,x^a), \qquad \epsilon=\epsilon(n_B,x^a),
\end{equation}
where \(n_B\) is the baryon density and \(x^a\) denotes the set of
independent composition variables. The frozen-composition sound speed is
defined as
\begin{equation}
c_s^2
=
\left(\frac{\partial P}{\partial \epsilon}\right)_{x^a}
=
\frac{P_n}{\epsilon_n},
\end{equation}
with
\begin{equation}
P_n
=
\left(\frac{\partial P}{\partial n_B}\right)_{x^a},
\qquad
\epsilon_n
=
\left(\frac{\partial \epsilon}{\partial n_B}\right)_{x^a}.
\end{equation}

Along the beta-equilibrated sequence, the composition variables become
functions of density, \(x^a=x_\beta^a(n_B)\). Therefore,
\begin{equation}
\frac{dP_\beta}{dn_B}
=
P_n+\sum_a P_a\frac{dx_\beta^a}{dn_B},
\end{equation}
and
\begin{equation}
\frac{d\epsilon_\beta}{dn_B}
=
\epsilon_n+\sum_a \epsilon_a\frac{dx_\beta^a}{dn_B},
\end{equation}
where
\begin{equation}
P_a
=
\left(\frac{\partial P}{\partial x^a}\right)_{n_B,x^{b\neq a}},
\qquad
\epsilon_a
=
\left(\frac{\partial \epsilon}{\partial x^a}\right)_{n_B,x^{b\neq a}} .
\end{equation}
The beta-equilibrium sound speed is then
\begin{equation}
c_e^2
=
\frac{dP_\beta/dn_B}{d\epsilon_\beta/dn_B}.
\end{equation}

The Ledoux buoyancy term entering the Brunt--V\"ais\"al\"a frequency is
defined as
\begin{equation}
\mathcal{L}
=
\frac{1}{c_e^2}
-
\frac{1}{c_s^2}.
\end{equation}
Using the chain rule, this quantity can be decomposed exactly into
channel-resolved contributions,
\begin{equation}
\mathcal{L}
=
\sum_a \mathcal{L}_a,
\end{equation}
with
\begin{equation}
\mathcal{L}_a
=
\frac{
\left(c_s^2\epsilon_a-P_a\right)
\dfrac{dx_\beta^a}{dn_B}
}{
c_s^2\,dP_\beta/dn_B
}.
\end{equation}
This identity provides a direct numerical check of the decomposition:
the reconstructed sum \(\sum_a \mathcal{L}_a\) must reproduce the total
Ledoux term \(\mathcal{L}\).

For the antikaon-condensed sequence, we choose
\begin{equation}
x^a=\{Y_e,Y_\mu,Y_{K^-}\}.
\end{equation}
The proton and neutron fractions are reconstructed from charge neutrality
and baryon-number conservation,
\begin{equation}
Y_p=Y_e+Y_\mu+Y_{K^-},
\end{equation}
and
\begin{equation}
Y_n=1-Y_p.
\end{equation}
Thus, \(\mathcal{L}_{K^-}\) denotes the charge-neutral \(K^-\) channel,
including the associated rearrangement of \(Y_p\) and \(Y_n\). The
per-channel decomposition is therefore defined with respect to this
chosen independent basis. The total Ledoux term \(\mathcal{L}\) is basis
independent, while the attribution of the constrained proton and neutron
rearrangement belongs to the chosen charge-neutral channel.

For the \(\Delta\)-admixed and hyperonic sequences, the same construction
is used. The independent variables are chosen as the lepton fractions and
the fractions of the non-nucleonic species, while \(Y_p\) and \(Y_n\) are
again reconstructed from charge neutrality and baryon-number conservation.

The channel-resolved Ledoux terms are mapped to the stellar background
using the metric convention
\begin{equation}
ds^2
=
-e^{\nu(r)}dt^2
+
e^{\lambda(r)}dr^2
+
r^2d\Omega^2 .
\end{equation}
We define the local gravitational acceleration as
\begin{equation}
g
=
\frac{1}{2}\frac{d\nu}{dr}.
\end{equation}
The Brunt--V\"ais\"al\"a frequency is written as
\begin{equation}
N^2(r)
=
g^2 e^{\nu-\lambda}\mathcal{L}[n_B(r)] .
\end{equation}
The species-resolved contributions are obtained by replacing
\(\mathcal{L}\) with \(\mathcal{L}_a\),
\begin{equation}
N_a^2(r)
=
g^2 e^{\nu-\lambda}\mathcal{L}_a[n_B(r)] ,
\end{equation}
so that, in the exact continuum Ledoux decomposition,
\begin{equation}
N^2(r)=\sum_a N_a^2(r).
\end{equation}

After solving the full-GR \(g_1\)-mode problem, we define the
positive-definite inertia-like radial weight
\begin{equation}
\begin{aligned}
\mathcal W_{g_1}(r)
={}&
(\varepsilon+P)e^{(\lambda-\nu)/2}r^{2l}
\\
&\times\left[
(\operatorname{Re}W)^2
+l(l+1)(\operatorname{Re}V)^2
\right].
\end{aligned}
\end{equation}
Here \(W\) and \(V\) are the full-GR Lindblom--Detweiler eigenfunctions.
Their arbitrary common phase is fixed by the normalization \(W(0)=1\),
and the implementation uses their real parts in \(\mathcal W_{g_1}\).
This is an inertia-like diagnostic weight, not the relativistic
canonical mode energy.

We define the signed channel and total integrals by
\begin{align}
I_a
&=
\int_0^R N_a^2(r)\mathcal W_{g_1}(r)\,dr,
\\
I_{\rm tot}
&=
\int_0^R N^2(r)\mathcal W_{g_1}(r)\,dr,
\end{align}
and
\begin{equation}
\chi_a=\frac{I_a}{I_{\rm tot}}.
\end{equation}
The total \(N^2\) in \(I_{\rm tot}\) is calculated independently from
the basis-independent total Ledoux term \(\mathcal L\), rather than
being replaced by a forced sum of the species integrals. A common
rescaling of \(W\) and \(V\) therefore cancels from \(\chi_a\).

Composition derivatives at particle thresholds are evaluated using
their one-sided continuum limits before radial quadrature. Since
\(N^2=\sum_a N_a^2\) in the continuum decomposition,
\(\sum_a\chi_a=1\); individual \(\chi_a\) may nevertheless lie outside
\([0,1]\) because the contributions are signed. Positive (negative)
\(\chi_a\) denotes a channel that supports (opposes) the net buoyant
restoring force sampled by the mode.

\section{Results}\label{res}

\begin{figure*}[htbp]
    \centering
    \includegraphics[width=0.98\textwidth]{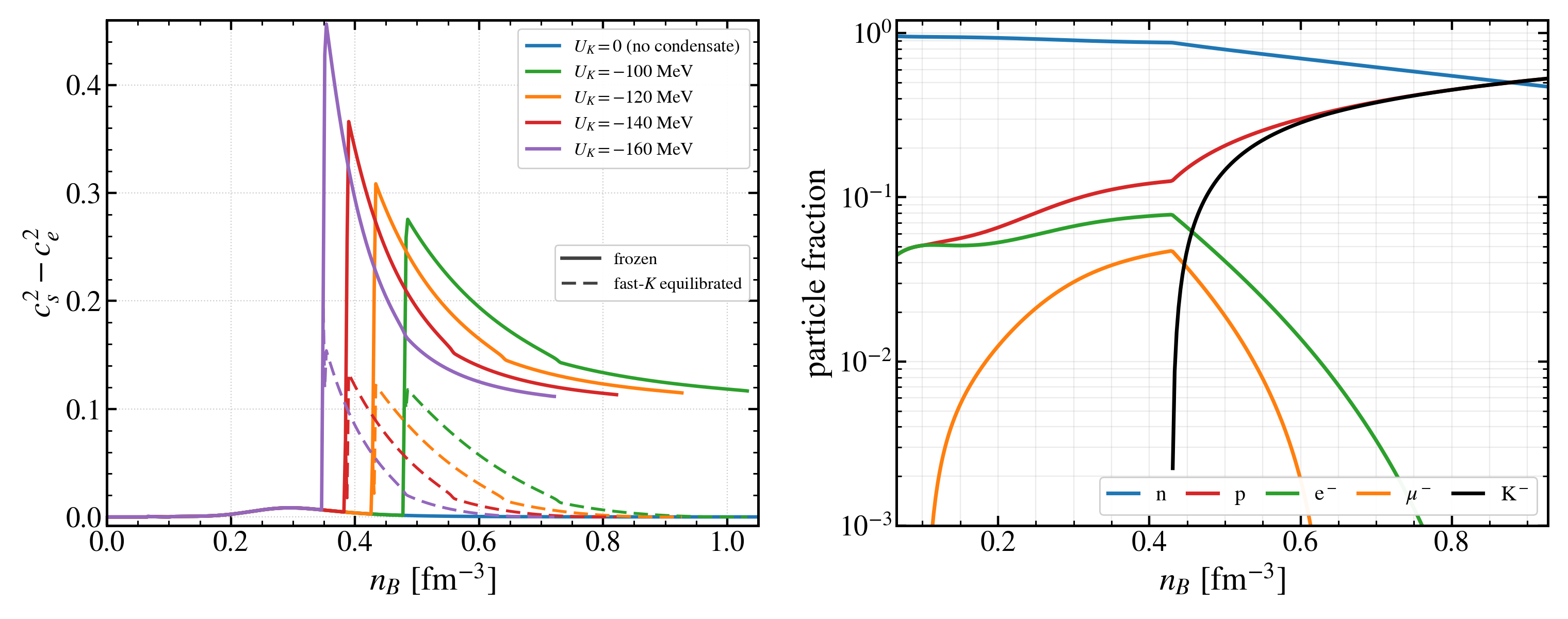}

   \caption{Difference between the adiabatic and equilibrium sound speeds,
$c_s^2-c_e^2$, as a function of baryon density $n_B$ for neutron-star
matter with antikaon condensation. Different colors correspond to
different antikaon optical potentials $U_K$, with $U_K=0$ MeV denoting
the purely nucleonic case without a condensate. Solid curves correspond
to fully frozen composition, whereas dashed curves show the fast-$K$
limit, in which $\delta Y_e=\delta Y_\mu=0$ and
$\delta(\mu_n-\mu_p-\mu_{K^-})=0$. The right panel shows the particle
fractions of neutrons, protons, electrons, muons, and condensed $K^-$
mesons for $U_K=-120$ MeV.}
    \label{fig:kaon_buoyancy}
\end{figure*}

\cref{fig:kaon_buoyancy} shows the density dependence of
$c_s^2-c_e^2$, which measures the local composition-induced buoyancy
of the stellar core. Here $c_e^2=dP/d\varepsilon$ is obtained along
the $\beta$-equilibrated EOS, whereas $c_s^2$ is evaluated under two
limiting composition constraints during the oscillation. The solid
curves correspond to fully frozen particle fractions, while the dashed
curves show the fast-$K$ limit, in which
$\delta Y_e=\delta Y_\mu=0$ and
$\delta(\mu_n-\mu_p-\mu_{K^-})=0$. Therefore,
$c_s^2=c_e^2$ corresponds to locally barotropic matter with no
composition-buoyancy restoring force, while $c_s^2>c_e^2$ gives stable
stratification and supports core $g$ modes.

For the purely nucleonic sequence, $c_s^2-c_e^2$ remains very small
over the full density range, indicating weak composition buoyancy in
the absence of an additional phase transition. Once the antikaon
condensate appears, the difference increases sharply at the onset
density. The onset is shifted to lower density for more attractive
antikaon optical potentials. The onset densities are
$n_B\simeq0.482$, $0.431$, $0.387$, and $0.351~{\rm fm}^{-3}$ for
$U_K=-100$, $-120$, $-140$, and $-160$ MeV, respectively. The frozen
and fast-$K$ results coincide below the condensation threshold, as
required. Above the threshold, fast-$K$ equilibration substantially
reduces the peak in $c_s^2-c_e^2$, whose maximum retains approximately
$36\%$--$44\%$ of its fully frozen value, but does not eliminate it.
The peak identifies the density layer where the kaon-induced
composition gradient is strongest. Beyond the peak, the condensate is
already established and the composition varies more smoothly with
density, causing $c_s^2-c_e^2$ to decrease. The decrease is stronger
in the fast-$K$ limit, for which the surviving buoyancy is concentrated
more closely around the condensate-onset layer.

The particle fractions for the representative case $U_K=-120$ MeV,
shown in the right panel, clarify the origin of this behaviour. Below
the condensation threshold, charge neutrality is maintained by the
electrons and muons, such that $Y_p=Y_e+Y_\mu$, while the proton and
lepton fractions increase gradually with density. At
$n_B\simeq0.431~{\rm fm}^{-3}$, the $K^-$ condensate appears and its
fraction increases rapidly. The condensed kaons then progressively
replace the leptons as the dominant carriers of negative charge, and
charge neutrality becomes
$Y_p=Y_e+Y_\mu+Y_{K^-}$. Consequently, both $Y_e$ and $Y_\mu$ decrease
above the threshold, with the muon fraction becoming strongly
suppressed at a lower density than the electron fraction. At the same
time, the negative charge supplied by the condensate permits a rapid
increase in the proton fraction, accompanied by a corresponding
decrease in the neutron fraction. At high density, where the lepton
fractions are small, $Y_p$ closely follows $Y_{K^-}$, as expected from
charge neutrality. The abrupt changes in these particle fractions at
the onset of condensation produce the sharp peak in $c_s^2-c_e^2$,
whereas their smoother evolution within the established kaon-condensed
phase accounts for the gradually decreasing post-onset tail.

\begin{figure}[t]
    \centering
        \centering
        \includegraphics[width=0.90\linewidth]{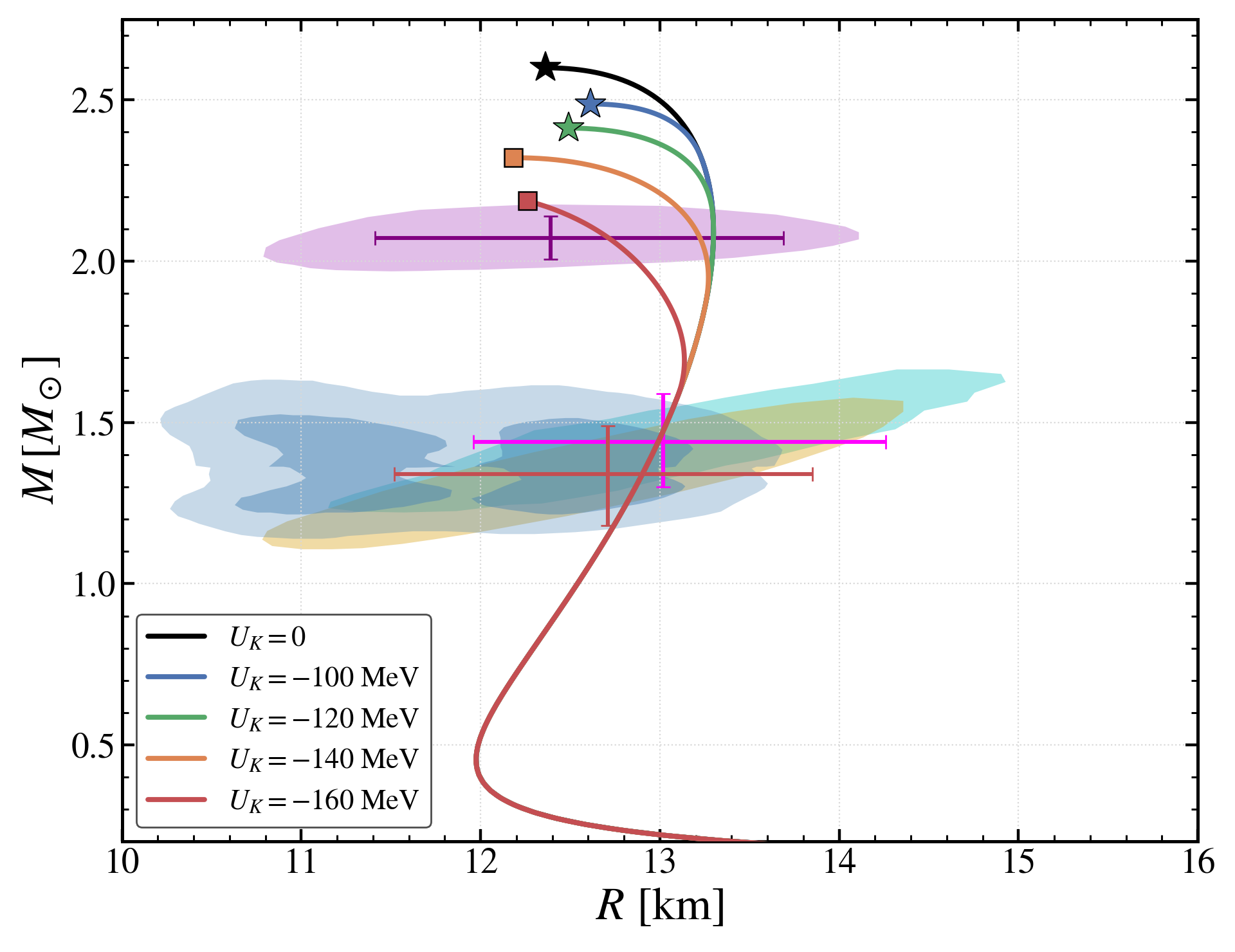}
        \label{fig:kaon_MR}
    \caption{Mass--radius relations for antikaon-condensed neutron-star
matter with different optical potentials $U_K$. The case $U_K=0$ MeV
denotes the purely nucleonic sequence without antikaon condensation,
Stars mark physical maximum-mass configurations, while squares mark
EOS-validity terminal configurations reached before a physical maximum. The
steel-blue contours show the $90\%$ and $50\%$ credible regions for
the two components of GW170817. The cyan and yellow regions show the
$68\%$ credible regions of the two-dimensional mass--radius posteriors
for PSR J0030+0451 \cite{Riley:2019yda,Miller:2019cac}, while the purple
high-mass region shows the corresponding $68\%$ credible regions for
PSR J0740+6620 \cite{Riley:2021pdl,Miller:2021qha}. }
    \label{fig:kaon_mass_radius}
\end{figure}

\begin{figure*}[htbp]
    \centering
    \includegraphics[width=0.48\textwidth]{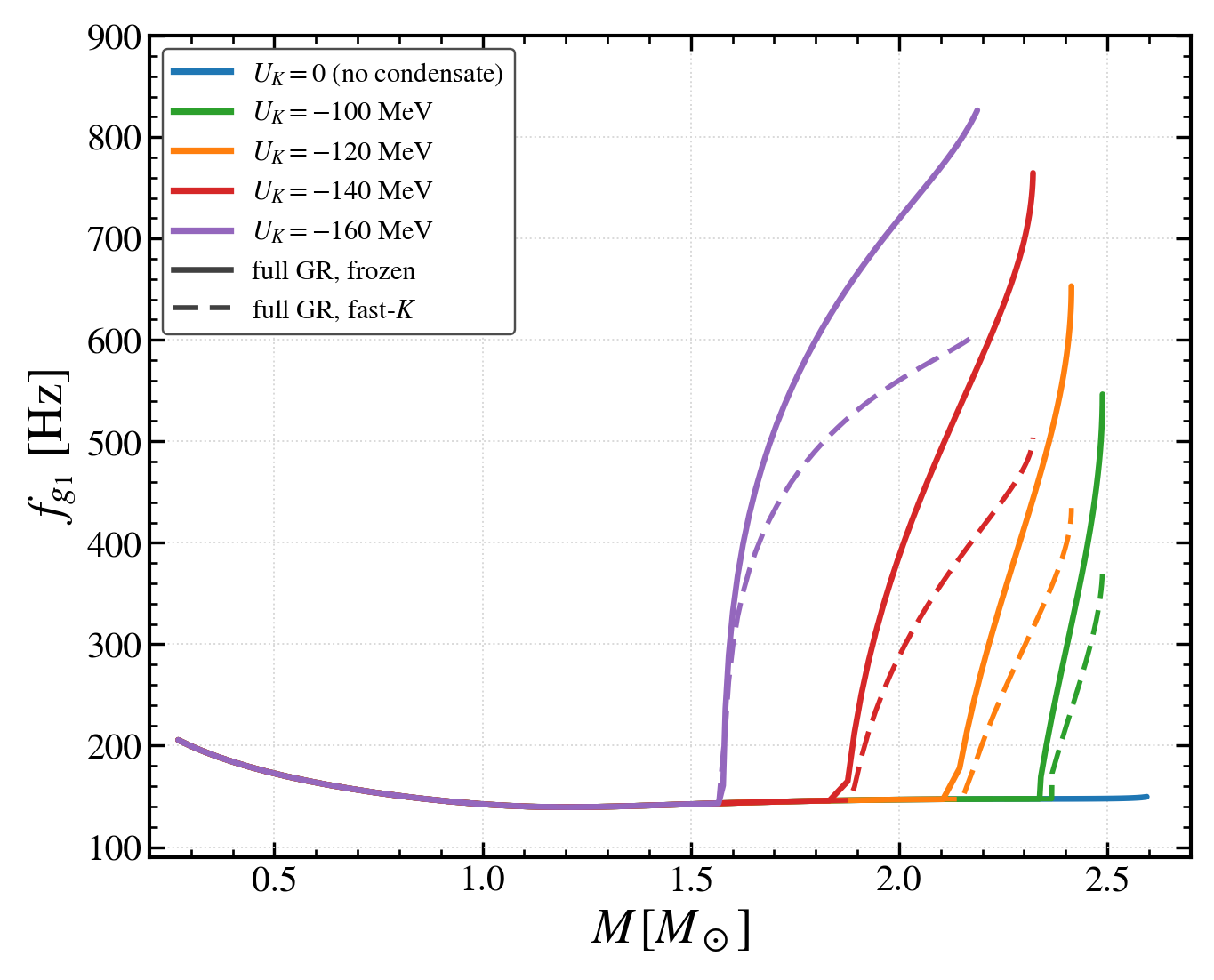}
    \hfill
    \includegraphics[width=0.48\textwidth]{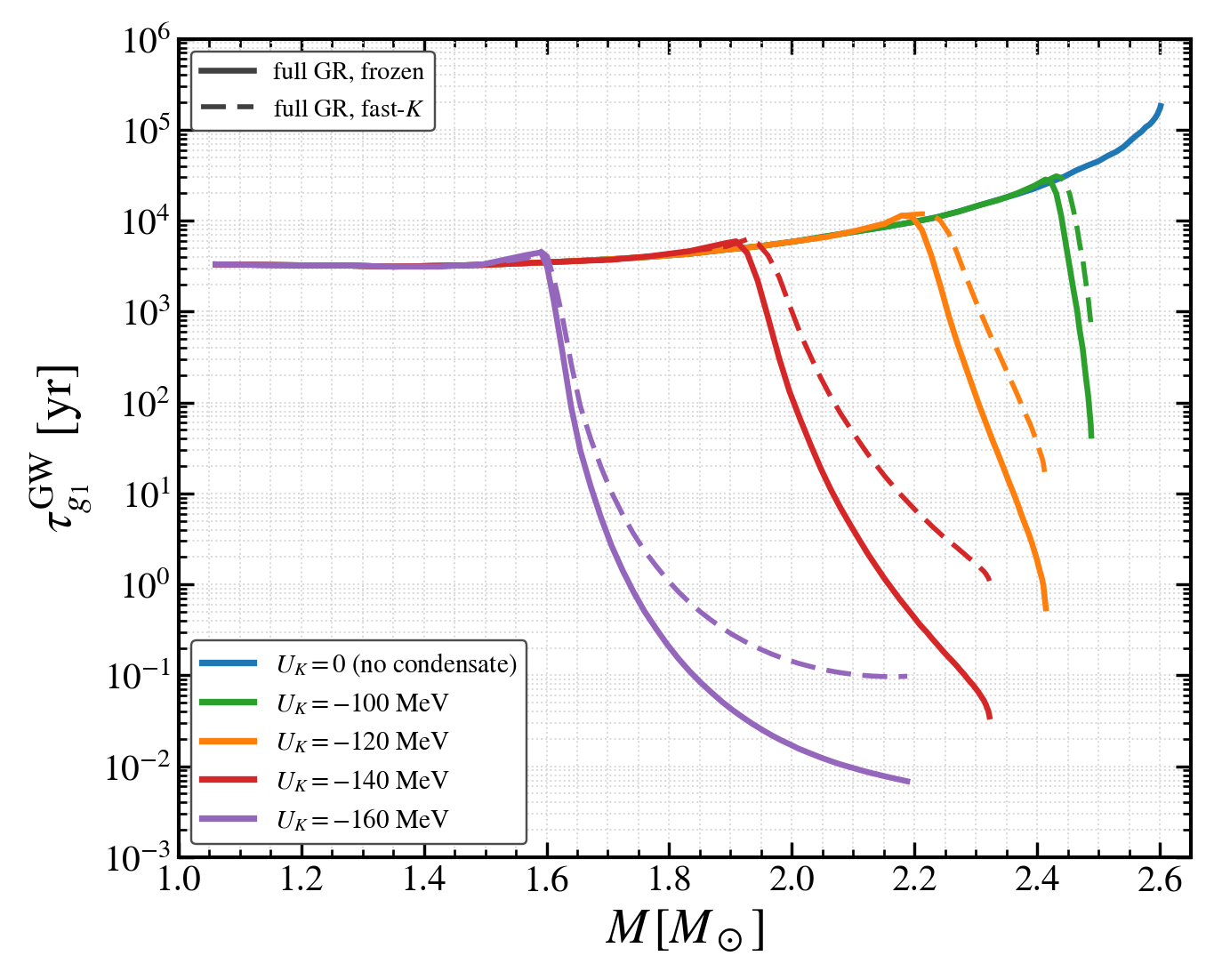}

    \caption{Full-GR $g_1$-mode frequencies (left) and gravitational-wave
damping times (right) as functions of gravitational mass for different
antikaon optical potentials $U_K$. The case $U_K=0$ MeV denotes the
purely nucleonic sequence without antikaon condensation. Solid curves
correspond to fully frozen composition, whereas dashed curves show the
fast-$K$ limit, in which $\delta Y_e=\delta Y_\mu=0$ and
$\delta(\mu_n-\mu_p-\mu_{K^-})=0$. The two limits coincide below the
kaon-condensation threshold.}
    \label{fig:kaon_gmode}
\end{figure*}

\cref{fig:kaon_mass_radius} shows the mass--radius sequences for
the different antikaon optical potentials together with the
observational mass--radius constraints. The steel-blue contours
represent the two components of GW170817, with the outer and inner
contours denoting the $90\%$ and $50\%$ credible regions, respectively.
The cyan and yellow regions show the $68\%$ credible regions of the
two-dimensional mass--radius posteriors for PSR J0030+0451
\cite{Riley:2019yda,Miller:2019cac}, while the purple high-mass region
shows the corresponding $68\%$ credible regions for PSR J0740+6620
\cite{Riley:2021pdl,Miller:2021qha}. All kaonic
sequences coincide with the purely nucleonic result at low and
canonical masses because their central densities remain below the
condensation threshold. Consequently, the canonical radius is
independent of $U_K$, with
$R_{1.4}=12.957$ km for all five sequences. The common canonical-mass
branch passes through the lower-mass observational regions displayed
in the figure, showing that the present constraints cannot distinguish
between these optical potentials before the condensate appears.

Once the central density exceeds the kaon-condensation threshold, the
mass--radius curves separate. Increasing the attraction of the optical
potential shifts the onset to lower stellar mass and produces stronger
high-density softening, making the massive configurations progressively
more compact. As summarized in \cref{tab:eos_summary_kaon},
the physical maximum masses are $2.600$, $2.487$, and
$2.413\,M_\odot$ for $U_K=0$, $-100$, and $-120$ MeV, respectively.
For $U_K=-140$ and $-160$ MeV, the sequences instead reach
EOS-validity terminal configurations at $2.321$ and $2.187\,M_\odot$
while the stellar mass is still rising, so these values are not
physical maxima. Every sequence nevertheless reaches above
$2\,M_\odot$ within the valid EOS domain, and the high-mass branches
remain compatible with the NICER mass--radius region for
PSR J0740+6620. Thus, the displayed
mass--radius constraints permit all four antikaon potentials and do
not by themselves resolve the high-density composition; their main
differences emerge only after condensation, motivating the use of
composition-sensitive $g$ modes.

The left panel of \cref{fig:kaon_gmode} shows the full-GR core
$g_1$-mode frequency as a function of gravitational mass. The purely
nucleonic mode remains on a low-frequency branch; for
$M\gtrsim1\,M_\odot$ it stays within approximately $139$--$151$ Hz and
reaches $150.66$ Hz at the physical maximum. This weak variation
reflects the small composition stratification of the nucleonic core.
Before the appearance of the condensate, all antikaon sequences follow
the same branch because their compositions and buoyancy profiles are
identical to those of the nucleonic sequence.

Once the $K^-$ condensate appears, the frozen-composition frequency
rises sharply above the nucleonic value, with the onset moving to lower
stellar mass as the optical potential deepens ($U_K=-160$ MeV first,
then $-140$, $-120$, and $-100$ MeV). At the terminal configurations, the
frozen frequency climbs monotonically with attraction, from $546$ Hz at
$U_K=-100$ MeV to $827$ Hz at $U_K=-160$ MeV (\cref{tab:eos_summary_kaon}
lists the full set); the two weakest potentials reach a physical maximum
mass, whereas the two strongest terminate at the EOS-validity limit.
Below the condensate onset, all sequences instead share the leptonic
value $f_{g_1}^{1.4}=140.97$ Hz at $1.4\,M_\odot$.

Fast-$K$ equilibration reduces this enhancement but does not eliminate
it. At the same terminal configurations (\cref{tab:eos_summary_kaon}), 
the fast-$K$ modes retain $65.7\%$--$73.4\%$ of their
frozen-composition frequencies and remain distinctly above the
nucleonic band. The surviving mode is consistent with the residual
post-onset buoyancy seen in \cref{fig:kaon_buoyancy}: allowing the
condensate to re-equilibrate removes part of the composition contrast
of a displaced fluid element, but the matter does not become locally
barotropic when the lepton fractions remain frozen. The frozen and
fast-$K$ calculations therefore provide two limiting responses of the
kaon-condensed core rather than qualitatively different mode branches.

The right panel of \cref{fig:kaon_gmode} shows the corresponding
full-GR gravitational-wave damping times. Below the condensation
threshold, the frozen and fast-$K$ results coincide. At canonical mass,
all sequences have $\tau_{g_1}^{\rm GW}=3188$ yr and follow the
nucleonic trend. Near the condensate threshold, the resolved curves
show a smooth, family-dependent feature on the $10^3$-yr scale before
turning downward. This feature reflects the rapid reorganization of
the mode as the kaon-induced buoyancy cavity first develops and remains
localized near the condensate-onset layer.

After the condensate occupies a finite part of the core,
$\tau_{g_1}^{\rm GW}$ decreases rapidly toward higher mass. In the frozen limit, the terminal damping times decrease from
\(24.26\) yr for \(U_K=-100\) MeV to
\(6.8\times10^{-3}\) yr for \(U_K=-160\) MeV
(\cref{tab:eos_summary_kaon}), and fast-\(K\) equilibration
lengthens them by factors of \(14.4\)--\(31.8\). This weaker gravitational-wave
emission is consistent with the reduced mode frequency and the
accompanying change in the mode eigenfunction when part of the
kaon-induced buoyancy is removed.
The computed roots satisfy \(\operatorname{Im}\omega<0\) under the
adopted \(e^{-i\omega t}\) convention, and hence neither treatment
produces a mode instability.

Even the shortest resolved damping time,
$6.8\times10^{-3}$ yr, remains many orders of magnitude longer than the
sub-second resonance-crossing time of a binary-neutron-star inspiral.
Gravitational-wave damping alone therefore does not suppress resonant
excitation of the kaon-induced $g_1$ mode. The damping times reported
here include only gravitational-wave emission. For a finite kaon
reaction rate, the mode frequency is expected to interpolate between
the two limiting responses, while additional chemical dissipation may
also arise. Determining the direct observability of the mode requires
the tidal-overlap integral and the accumulated gravitational-wave phase
shift, considered separately below.

\begin{table*}[htbp]
\centering
\caption{Stellar properties and full-GR $g_1$-mode results for the
BigApple$+K^-$ EOS family. The frozen and fast-$K$ quantities are
evaluated on the same terminal TOV configuration: the maximum-mass
model for $U_K=0,-100,-120$ and the EOS-validity terminal model
(daggered) for $U_K=-140,-160$. At $1.4\,M_\odot$, all sequences
remain below the kaon-condensation threshold and therefore share the
frozen-composition result. A dash indicates that the fast-$K$ limit is
not applicable to the purely nucleonic sequence.}
\label{tab:eos_summary_kaon}

\begingroup
\small
\setlength{\tabcolsep}{4.2pt}
\renewcommand{\arraystretch}{1.3}

\begin{tabular}{c c c c c c c c c c}
\hline\hline
& \multicolumn{3}{c}{Stellar properties}
& \multicolumn{3}{c}{$f_{g_1}$ (Hz)}
& \multicolumn{3}{c}{$\tau_{g_1}^{\rm GW}$ (yr)} \\
\cline{2-4}\cline{5-7}\cline{8-10}
$U_K$ &
$M_{\rm term}$ &
$R_{\rm term}$ &
$R_{1.4}$ &
$f_{g_1}^{\rm term,frozen}$ &
$f_{g_1}^{\rm term,\text{fast-}K}$ &
$f_{g_1}^{1.4}$ &
$\tau_{g_1}^{\rm term,frozen}$ &
$\tau_{g_1}^{\rm term,\text{fast-}K}$ &
$\tau_{g_1}^{1.4}$ \\
(MeV) &
($M_\odot$) &
(km) &
(km) &
&
&
&
&
&
\\
\hline
$0$    & 2.600 & 12.362 & 12.957 & 150.66 & --     & 140.97 & $1.84\times10^{5}$ & --      & 3188 \\
$-100$ & 2.487 & 12.618 & 12.957 & 545.95 & 377.34 & 140.97 & 24.26   & 372.1   & 3188 \\
$-120$ & 2.413 & 12.484 & 12.957 & 651.69 & 433.75 & 140.97 & 0.492   & 12.60   & 3188 \\
$-140$ & 2.321$^{\dagger}$ & 12.184 & 12.957 & 770.43 & 506.25 & 140.97 & 0.033   & 1.051   & 3188 \\
$-160$ & 2.187$^{\dagger}$ & 12.261 & 12.957 & 826.53 & 606.73 & 140.97 & 0.0068  & 0.0978  & 3188 \\
\hline\hline
\end{tabular}

\endgroup

\vspace{2pt}
{\footnotesize $^{\dagger}$EOS-validity endpoint: the mass sequence is still
rising when the EOS reaches its validity limit, before a physical maximum
is reached.}
\end{table*}

\begin{table}[t]
\centering
\caption{Species-resolved Ledoux decomposition of the \(g_1\)-mode
using the inertia-like weight \(\mathcal W_{g_1}\) for the
antikaon-condensed sequence (\(U_K=-120\) MeV). The listed \(\chi_a\)
values use fully frozen composition. \(n_{B,c}\) is the central baryon
density, and \(\chi_a\) is the signed contribution of channel \(a\) to
the buoyant driving sampled by the full-GR \(g_1\)-mode eigenfunction.}
\label{tab:kaon_chi}
\begin{ruledtabular}
\begin{tabular}{ccccc}
\(M/M_\odot\) & \(n_{B,c}\,(\mathrm{fm}^{-3})\) & \(f_{g_1}\,(\mathrm{Hz})\) &
\(\chi_{K^-}\) & \(\chi_{e+\mu}\) \\
\hline
1.999 & 0.410 & 146.7 & 0.000 & $1.058$ \\
2.199 & 0.473 & 277.0 & 1.046 & $-0.046$ \\
2.299 & 0.534 & 414.5 & 1.024 & $-0.024$ \\
\end{tabular}
\end{ruledtabular}
\end{table}

Applying the species-resolved decomposition to the representative
\(U_K=-120\) MeV sequence (\cref{tab:kaon_chi}) shows that the antikaon
\(g_1\)-mode is threshold-activated: it becomes kaon-dominated only once
the stellar core crosses the condensate onset,
\(n_B^{\rm onset}\simeq0.431~\mathrm{fm}^{-3}\). Just below onset
(at \(2.0\,M_\odot\), central density \(0.410~\mathrm{fm}^{-3}\)) the
mode is the ordinary lepton-driven composition mode, \(\chi_{K^-}=0\);
once the core exceeds the threshold \(\chi_{K^-}\) jumps to near unity
and stays there, so the charge-neutral \(K^-\) Ledoux layer dominates
the buoyant restoring force at higher mass. Values of \(\chi_{K^-}\)
slightly above unity reflect the signed nature of the decomposition.
The antikaon-induced \(g_1\)-mode is therefore a high-mass,
threshold-activated composition mode, absent at canonical mass.

\subsection*{Hyperons and $\Delta$ resonances}

\begin{figure*}[htbp]
    \centering
    \includegraphics[width=0.98\textwidth]{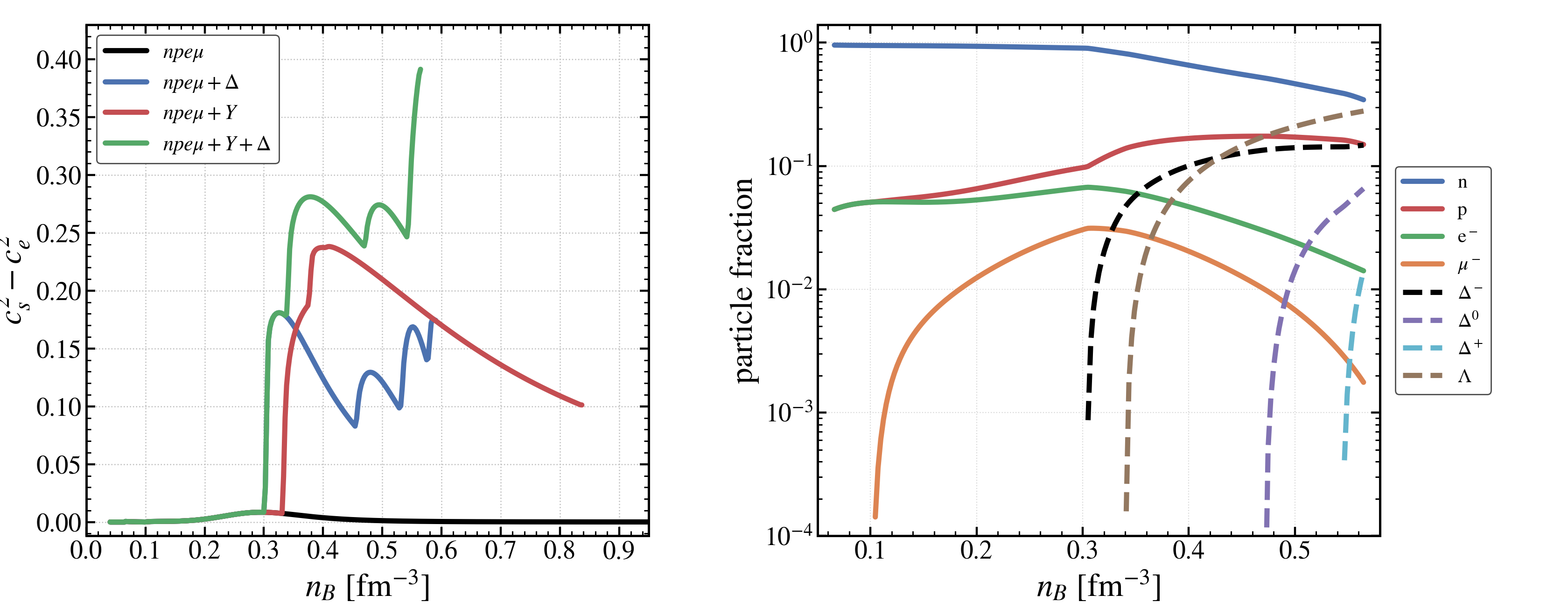}

    \caption{Left: Difference between the frozen-composition and
equilibrium sound speeds, $c_s^2-c_e^2$, as a function of baryon
density $n_B$ for the $npe\mu$, $npe\mu+\Delta$, $npe\mu+Y$, and
$npe\mu+Y+\Delta$ compositions. Right: Equilibrium particle fractions
for the representative $npe\mu+Y+\Delta$ sequence, showing the
successive appearance of $\Delta^-$, $\Lambda$, $\Delta^0$, and
$\Delta^+$.}
    \label{fig:hd_buoyancy}
\end{figure*}

The inclusion of hyperons and \(\Delta\) baryons changes the buoyancy
profile qualitatively. The left panel of \cref{fig:hd_buoyancy} compares
\(c_s^2-c_e^2\) for the nucleonic \(npe\mu\) sequence, the
\(\Delta\)-admixed \(npe\mu+\Delta\) sequence, the hyperonic
\(npe\mu+Y\) sequence, and matter containing both hyperons and
\(\Delta\) resonances. The nucleonic contribution remains weak, whereas
the appearance of non-nucleonic degrees of freedom produces pronounced
and extended buoyancy structures. In the \(npe\mu+\Delta\) sequence,
the \(\Delta^-\) appears first at
\(n_B\simeq0.305~\mathrm{fm}^{-3}\) and generates the first major
enhancement. The subsequent onsets of \(\Delta^0\), \(\Delta^+\), and
\(\Delta^{++}\), at approximately \(0.459\), \(0.534\), and
\(0.580~\mathrm{fm}^{-3}\), respectively, produce the additional
high-density structure.

The particle fractions shown in the right panel clarify the corresponding
behavior of the \(npe\mu+Y+\Delta\) sequence. The \(\Delta^-\) again
appears first and rapidly replaces electrons and muons as a carrier of
negative charge, producing a strong rearrangement of the proton and
lepton fractions and the dominant rise in \(c_s^2-c_e^2\). The neutral
\(\Lambda\) hyperon appears next at
\(n_B\simeq0.341~\mathrm{fm}^{-3}\), followed at higher density by
\(\Delta^0\) and \(\Delta^+\), generating the subsequent features in the
buoyancy profile. The \(\Delta^{++}\) fraction remains marginal in this
sequence, while the \(\Sigma\) and \(\Xi\) hyperons are suppressed
because the appearance of \(\Delta^-\) strongly reduces the electron
chemical potential. Thus, the combined sequence contains three
appreciable \(\Delta\) charge states together with the \(\Lambda\).
Unlike antikaon condensation, which produces a localized threshold
feature, the successive baryonic onsets generate a broader,
multi-step buoyancy profile extending over a substantial part of the
stellar core.

\begin{figure}[t]
    \centering
\includegraphics[width=0.90\linewidth]{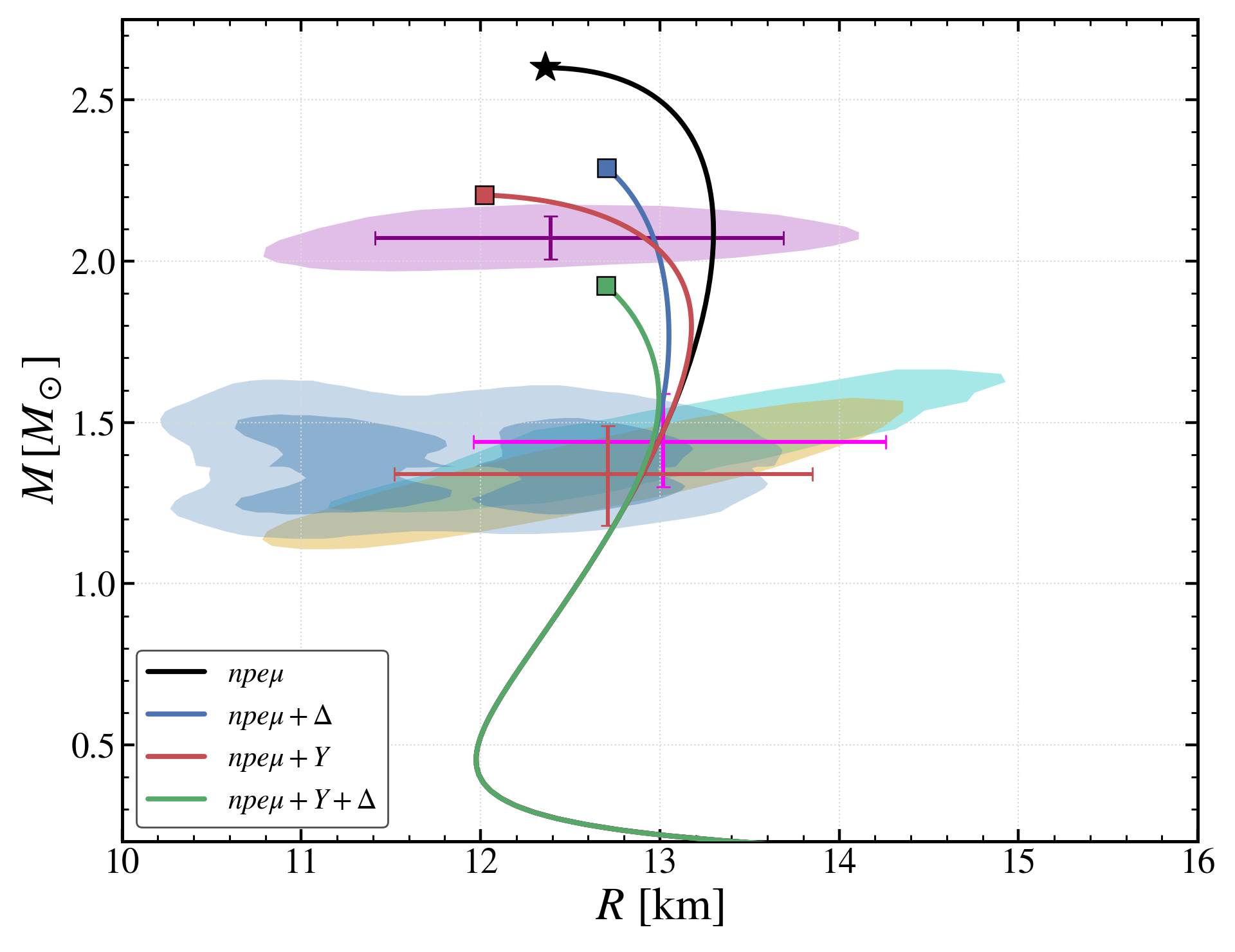}
    \caption{Mass--radius sequences for the $npe\mu$, $npe\mu+\Delta$,
$npe\mu+Y$, and $npe\mu+Y+\Delta$ compositions. The contour regions are the same as in \cref{fig:kaon_mass_radius}. The star marks the
physical maximum of the nucleonic sequence, while squares mark the
EOS-validity terminal configurations of the $N\Delta$, $NY$, and
$NY\Delta$ sequences, reached before physical maxima.}
    \label{fig:hd_mr}
\end{figure}

\begin{figure*}[t]
    \centering
    \begin{subfigure}[b]{0.45\textwidth}
        \centering
        \includegraphics[width=\textwidth]{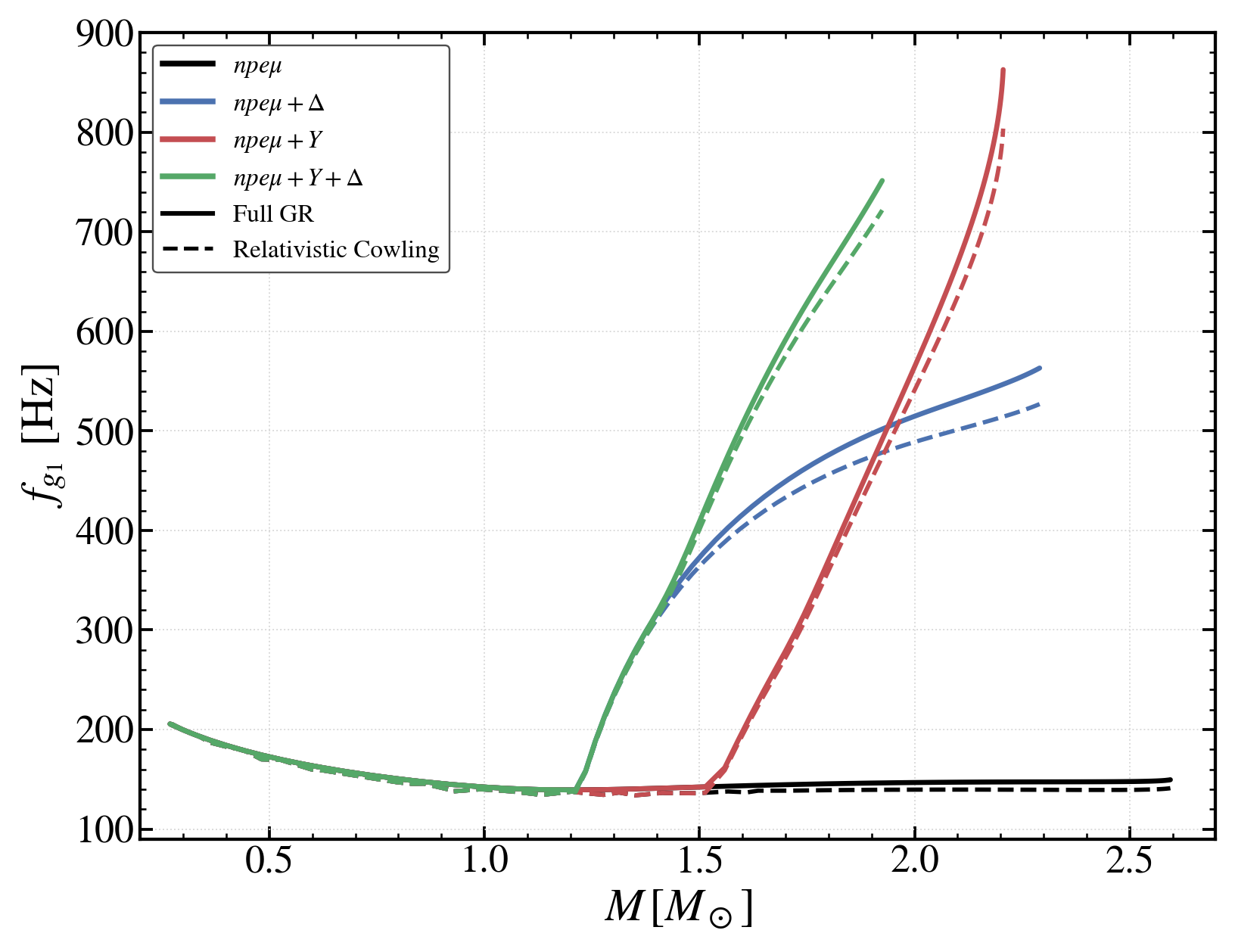}
        \label{fig:hd_g1}
    \end{subfigure}
    \hfill
    \begin{subfigure}[b]{0.45\textwidth}
        \centering
        \includegraphics[width=\textwidth]{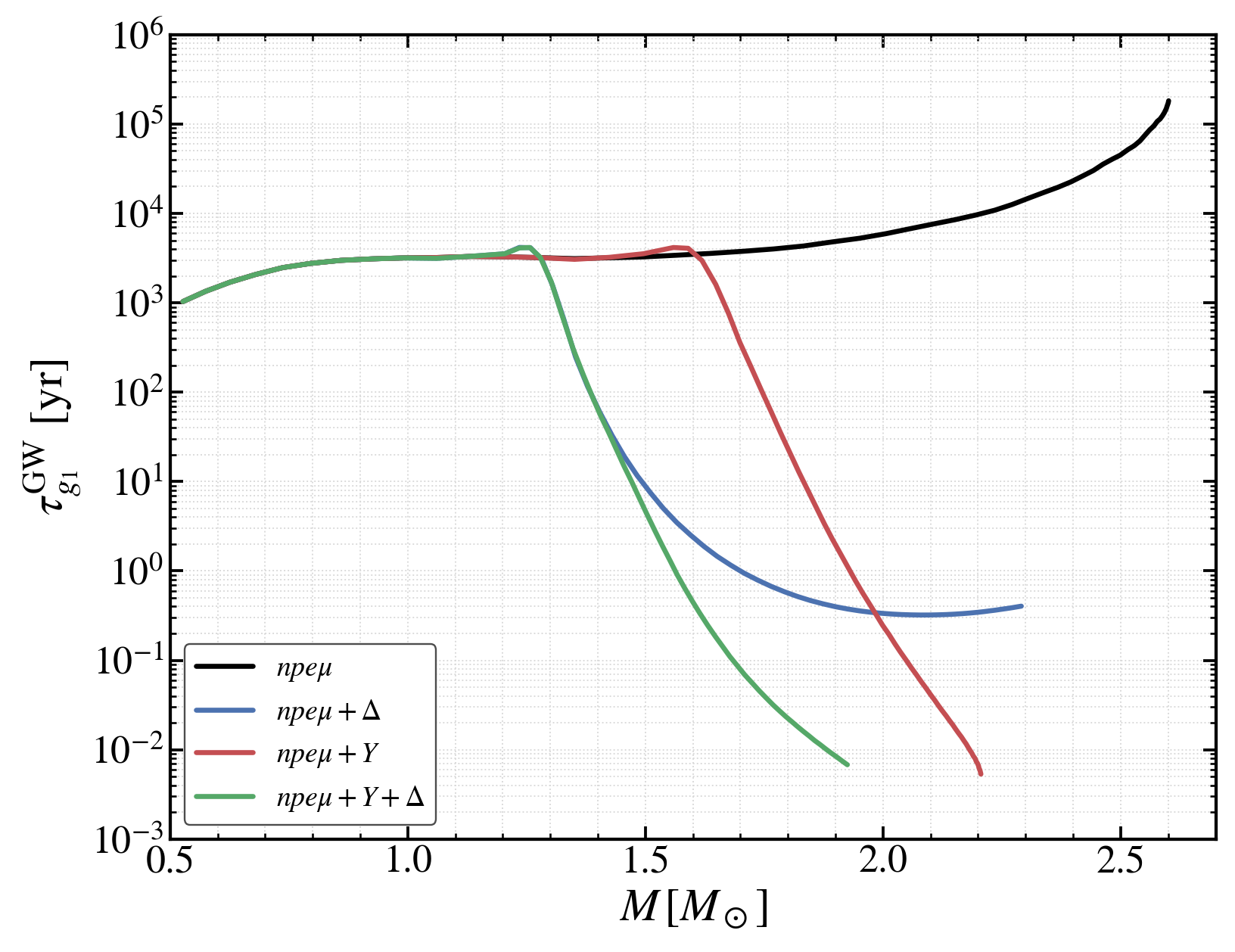}
        \label{fig:hd_damping}
    \end{subfigure}
\caption{Left: Full-GR core \(g_1\)-mode frequency as a function of stellar mass for the \(npe\mu\), \(npe\mu+\Delta\), \(npe\mu+Y\), and \(npe\mu+Y+\Delta\) sequences. Solid and dashed curves denote the full-GR and relativistic Cowling results, respectively. Right: Gravitational-wave damping time of the full-GR core \(g_1\)-mode for the same compositions.}
    \label{fig:delta_gmode}
\end{figure*}

\cref{fig:hd_mr} shows the mass--radius relations for the
\(npe\mu\), \(npe\mu+\Delta\), \(npe\mu+Y\), and
\(npe\mu+Y+\Delta\) sequences, while
\cref{tab:eos_summary} summarizes their stellar and oscillation
properties. Relative to the nucleonic sequence, the inclusion of
non-nucleonic baryons softens the EOS and lowers the largest mass reached
within the calculated domain. Only the nucleonic sequence reaches a
physical maximum, at \(2.600\,M_\odot\). The \(npe\mu+Y\),
\(npe\mu+\Delta\), and \(npe\mu+Y+\Delta\) sequences instead terminate
at the EOS-validity limits \(2.205\), \(2.290\), and
\(1.924\,M_\odot\), respectively, while the stellar mass is still
rising. These terminal masses are neither physical maxima nor lower
bounds on them. The effect on the canonical radius is small:
\(R_{1.4}=12.957\) km for the \(npe\mu\) and \(npe\mu+Y\) sequences,
compared with \(12.935\) km for both the \(npe\mu+\Delta\) and
\(npe\mu+Y+\Delta\) sequences. The mass--radius curves
are compared in \cref{fig:hd_mr} with the constraints from
GW170817 \cite{LIGOScientific:2017vwq} and the NICER measurements of
PSR~J0030+0451 and PSR~J0740+6620
\cite{Riley:2021pdl,Miller:2021qha}.

The left panel of \cref{fig:delta_gmode} shows that the core
\(g_1\)-mode is substantially more sensitive to the appearance of
non-nucleonic baryons than the canonical radius. Before the first exotic
species appears, all four sequences follow the nucleonic branch, with
the full-GR frequency near \(141\) Hz at \(1.4\,M_\odot\).
For the two \(\Delta\)-bearing sequences, the departure from this branch
begins at \(M\simeq1.18\,M_\odot\), following the onset of the
\(\Delta^-\). 
At \(1.4\,M_\odot\) the two \(\Delta\)-bearing sequences already sit
near \(316\) Hz, whereas the nucleonic and hyperon-only sequences remain
at \(140.97\) Hz: at canonical mass the enhancement comes from the
\(\Delta\) sector, not the hyperons. Toward higher mass each new charge
state strengthens the buoyancy, and the terminal frequencies climb to
several hundred Hz---highest for the hyperon-only sequence once the
\(\Lambda\) appears (\cref{tab:eos_summary} gives the values). This
mirrors the buoyancy profiles in \cref{fig:hd_buoyancy}: every new
composition gradient lowers the equilibrium sound speed and strengthens
the restoring force that sets the \(g_1\)-mode.
All numerical frequencies quoted here are full-GR
results. The relativistic Cowling curves follow the same
mass dependence and remain within approximately \(10\%\) of the full-GR
frequencies, in agreement with previous calculations of composition
\(g\)-modes \cite{Sun:2026agt,Zhao:2022toc}.

Full-GR eigenfunction tracking confirms that the plotted mode follows a
continuous \(g_1\) branch across the \(\Delta^-\) and \(\Lambda\) onset
regions, with no interchange with a neighboring mode. The frequency
increase therefore reflects the changing composition buoyancy rather
than a switch in mode identity.

The right panel of \cref{fig:delta_gmode} shows the corresponding
gravitational-wave damping times. At low mass, before the appearance of
exotic baryons, all sequences follow the same nucleonic trend. The onset
of a new species produces a localized feature in the damping time,
followed by a rapid decrease as the high-frequency exotic-baryon mode
develops. At \(1.4\,M_\odot\) the \(\Delta\)-bearing sequences already damp
faster (\(\tau_{g_1}^{\rm GW}\simeq64\) yr) than the nucleonic and
hyperon-only sequences (\(3188\) yr). At the terminal configurations,
the damping times are \(0.404\) yr for \(N\Delta\),
\(5.3\times10^{-3}\) yr for \(NY\), and
\(6.8\times10^{-3}\) yr for \(NY\Delta\)
(\cref{tab:eos_summary}). Only the
nucleonic value is evaluated at a physical maximum; the other three are
EOS-validity terminal configurations.
 All computed quasinormal-mode
roots satisfy
\(\operatorname{Im}\omega<0\) under the adopted \(e^{-i\omega t}\)
convention; therefore, the shorter damping times represent more
efficient coupling to spacetime perturbations and gravitational
radiation rather than a mode instability. These damping times remain
much longer than the resonance-crossing time during binary inspiral, so
gravitational-wave damping alone is not expected to suppress resonant
tidal excitation.

  \begin{table}[t]
  \centering
  \caption{Species-resolved Ledoux decomposition of the full-GR
  \(g_1\)-mode using the inertia-like weight \(\mathcal W_{g_1}\) for
  the \(\Delta\)-admixed and hyperon+\(\Delta\) sequences. The listed
  \(\chi_a\) values use fully frozen composition. Here
  \(N\equiv npe\mu\), \(\chi_\ell=\chi_e+\chi_\mu\),
  \(\chi_\Delta=\sum_i\chi_{\Delta_i}\), and
  \(\chi_Y=\chi_\Lambda+\chi_\Sigma+\chi_\Xi\).}
  \label{tab:hypdelta_chi}
  \begin{tabular}{lccccc}
  \hline\hline
  EOS & \(M/M_\odot\) & \(f_{g_1}\,(\mathrm{Hz})\) &
  \(\chi_{\Delta}\) & \(\chi_Y\) & \(\chi_\ell\) \\
  \hline
  \(N\Delta\)  & 1.399 & 315.8 & 0.952 & --    & 0.034 \\
  \(NY\Delta\) & 1.400 & 316.9 & 0.943 & 0.004 & 0.033 \\
  \(NY\Delta\) & 1.599 & 507.4 & 0.641 & 0.335 & 0.014 \\
  \(NY\Delta\) & 1.799 & 663.8 & 0.511 & 0.458 & 0.024 \\
  \hline\hline
  \end{tabular}
  \end{table}

Applying the same decomposition to the \(\Delta\)-admixed and
hyperon+\(\Delta\) sequences (\cref{tab:hypdelta_chi}) shows that the
canonical-mass \(g_1\)-mode is driven almost entirely by the
charge-neutral \(\Delta^-\) channel: at \(1.4\,M_\odot\) the grouped
\(\Delta\) contribution is \(\chi_\Delta\simeq0.95\) for both the
\(N\Delta\) and \(NY\Delta\) sequences, the leptons carry the small
remainder, and hyperons are still negligible (\(\chi_Y\simeq0\)). Toward
higher mass the driving shifts rapidly to the \(\Lambda\) channel, and
by \(1.8\,M_\odot\) the \(\Delta\) and hyperon contributions are
comparable (\cref{tab:hypdelta_chi}). A separate exact-mass calculation
at \(1.9\,M_\odot\) resolves the grouped \(\Delta\) into
\(\chi_{\Delta^-}=0.457\) and \(\chi_{\Delta^0}=0.013\) against
\(\chi_\Lambda=0.492\), i.e.\ near-equipartition of the \(\Delta\) and
hyperon driving with no meaningful dominance ordering. The \(\Delta^-\)
channel therefore controls the canonical-mass mode, whereas hyperons
become dynamically important along the high-mass branch; as shown in
Sec.~\ref{sec:bracket}, this attribution directly predicts which part of
the buoyancy survives strong-interaction re-equilibration.

As an independent validation of the species
attribution, we also calculate the full-GR frequency-sensitivity
coefficient
\[
C_a=\frac{\partial\ln\omega^2}{\partial\ln\alpha_a},
\qquad
\widehat{C}_a=\frac{C_a}{\sum_b C_b},
\]
where \(\alpha_a\) rescales only the Ledoux contribution
\(\mathcal{L}_a\). For the \(N\Delta\) sequence at
\(1.4\,M_\odot\), we obtain
\(\widehat{C}_{\Delta}=0.990\), identifying the same dominant channel
as \(\chi_\Delta=0.952\) at the canonical-mass Table-III point. For the
\(NY\Delta\) sequence at \(1.9\,M_\odot\), the individual signed
coefficients are \(\widehat{C}_{\Delta^-}=0.501\),
\(\widehat{C}_{\Delta^0}=0.0075\), and
\(\widehat{C}_{\Lambda}=0.496\), giving the grouped values
\(\widehat{C}_{\Delta}=0.5085\) and \(\widehat{C}_{Y}=0.496\).
The sensitivity and inertia-weighted diagnostics therefore identify the
same main channels and the same near-equipartition, although their small
numerical ordering is not identical. For the separate exact-mass
antikaon sensitivity calculation at \(2.4\,M_\odot\),
\(\widehat{C}_{K^-}=1.0245\) and \(\chi_{K^-}=0.999\); this comparison
does not use \cref{tab:kaon_chi}. The sensitivity coefficients
identify the same dominant buoyancy channels as the inertia-weighted
decomposition. The
subdominant lepton contribution need not agree quantitatively because
\(\chi_a\) measures buoyant driving weighted by the inertia-like
diagnostic \(\mathcal W_{g_1}\), whereas
\(\widehat{C}_a\) measures the response of the global eigenfrequency
to rescaling a local buoyancy component.

\begin{table*}[htbp]
  \centering
  \caption{Stellar properties for the BigApple EOS family with different
  particle compositions. Here ``term'' denotes the terminal configuration:
  for the purely nucleonic sequence it is the physical maximum-mass model;
  for the daggered \(NY\), \(N\Delta\), and \(NY\Delta\) sequences it is the
  last model allowed by the EOS-validity condition, before a physical maximum
  is reached.}
  \label{tab:eos_summary}
  \renewcommand{\arraystretch}{1.4}
  \begin{tabular}{c @{\hspace{1.5em}} c @{\hspace{1.5em}} c @{\hspace{1.5em}} c @{\hspace{1.5em}} c @{\hspace{1.5em}} c @{\hspace{1.5em}} c @{\hspace{1.5em}} c}
  \hline\hline

  \large $EoS$ &
  \large $M_{\mathrm{term}}$ &
  \large $R_{\mathrm{term}}$ &
  \large $R_{1.4}$ &
  \large $f_{g_1}^{\mathrm{term}}$ &
  \large $f_{g_1}^{1.4}$ &
  \large $\tau_{g_1}^{\mathrm{term}}$ &
  \large $\tau_{g_1}^{1.4}$ \\

  &
  $(M_\odot)$ &
  (km) &
  (km) &
  (Hz) &
  (Hz) &
  (yr) &
  (yr) \\
\hline\hline
  $npe\mu$
  & 2.600 & 12.362 & 12.957 & 150.66 & 140.97 & $1.84\times10^{5}$ & 3188 \\

  $npe\mu+\Delta$
  & 2.290$^{\dagger}$ & 12.702 & 12.935 & 563.21 & 316.16 & 0.404 & 65.2 \\
  
  $npe\mu+Y$
  & 2.205$^{\dagger}$ & 12.020 & 12.957 & 863.03 & 140.97 & 0.0053 & 3188 \\
  
  $npe\mu+Y+\Delta$ 
  & 1.924$^{\dagger}$ & 12.700 & 12.935 & 751.56 & 317.29 & 0.0068 & 63.8 \\
  
  \hline\hline
  \end{tabular}
  
  \vspace{2pt}
  {\footnotesize $^{\dagger}$EOS-validity endpoint: the mass sequence is still
  rising when the EOS reaches its validity limit, before a physical maximum
  is reached.}
  \end{table*}

\subsection{Reaction-timescale limits for antikaon and $\Delta$ buoyancy}

  \label{sec:bracket}
Composition \(g\)-modes are supported when at least one
independent particle fraction of a displaced fluid element cannot
follow the equilibrium composition of its surroundings within an
oscillation period. The standard calculation therefore evaluates the
adiabatic sound speed \(c_s^2\) at completely frozen composition
\cite{1992ApJ...395..240R,Lai:1993di,Tran:2022dva,Sun:2026agt}.
For a reaction channel with relaxation time \(\tau_{\rm rel}\), the
frozen and reaction-equilibrated limits correspond to
\(\omega\tau_{\rm rel}\gg1\) and
\(\omega\tau_{\rm rel}\ll1\), respectively. At intermediate reaction
rates, the response becomes complex and frequency dependent, modifying
the restoring force and introducing chemical damping
\cite{Counsell:2023pqp,Zhao:2025pgx}. For the nucleonic Urca processes
considered in representative finite-temperature calculations,
equilibration becomes important for \(g\)-modes mainly at MeV-scale
temperatures, supporting the frozen treatment for cold inspiralling
neutron stars \cite{Zhao:2025pgx}. The limiting calculations below
therefore characterize the slow- and fast-reaction responses but do not
replace a finite-rate kinetic calculation.

The antikaon sector requires a separate reaction-rate
test. Conversion between nucleons and the condensate can proceed through
the nonleptonic weak reaction
\(n\leftrightarrow p+K^-\), whose relaxation time is distinct from that
of the nucleonic Urca reactions \cite{Chatterjee:2007qs}. We therefore
calculate two limiting responses on the same stellar backgrounds. In
the fully frozen limit all particle fractions remain fixed, whereas in
the fast-\(K\) limit the lepton fractions remain frozen while the kaon
reaction maintains chemical equilibrium,
\[
\delta Y_e=\delta Y_\mu=0,
\qquad
\delta(\mu_n-\mu_p-\mu_{K^-})=0 .
\]
The two prescriptions coincide below the condensation threshold, where
\(Y_{K^-}=0\). For the cold inspiralling stars considered here, deciding between these
limits requires the in-medium kaon relaxation rate at the mode
frequency. Because that finite-rate calculation is beyond the present
scope, we retain both limits as a bracket and do not assign either one
a priori to a particular temperature regime. The resulting limits are compared locally in
\cref{fig:kaon_buoyancy} and globally in
\cref{fig:kaon_gmode}. Fast-\(K\) equilibration reduces the maximum
local value of \(c_s^2-c_e^2\) to approximately \(36\%\)--\(44\%\) of
its frozen value but does not eliminate the post-onset buoyancy.
Maintaining kaon chemical equilibrium does not make the matter
barotropic because the frozen lepton fractions, together with charge
neutrality, require compensating changes in the proton and neutron
fractions. At the terminal configurations, the fast-\(K\)
frequencies retain approximately \(65.7\%\)--\(73.4\%\) of the frozen
frequencies and remain distinctly above the nucleonic band, while the
gravitational-wave damping times increase by factors of approximately
\(14.4\)--\(31.8\). The kaon-associated high-frequency branch therefore
survives in both limiting treatments and is not solely an artifact of
completely frozen kaon composition. This conclusion is not
rate-independent: an intermediate reaction rate may further modify the
frequency and introduce chemical damping, which is not included in the
reported \(\tau_{g_1}^{\rm GW}\).

The \(\Delta\) sector provides a contrasting case.
The reaction \(nn\leftrightarrow p\Delta^-\), together with analogous
reactions involving the other charge states, conserves baryon number,
electric charge, and strangeness and can proceed through the strong
interaction.  The free-space width
\(\Gamma_\Delta\simeq117\) MeV
\cite{ParticleDataGroup:2024cfk} illustrates the underlying
strong-interaction scale, but it does not by itself determine the
chemical relaxation rate in degenerate stellar matter, where
phase-space restrictions, Pauli blocking, and in-medium effects may be
important. We therefore calculate both limiting responses: fully frozen
composition and strong equilibrium within the \(\Delta\) quartet at
fixed lepton fractions and strangeness. These correspond to the slow-
and fast-\(\Delta\) limits, respectively. Existing \(\Delta\)-mode
calculations generally employ the frozen-composition treatment
\cite{Sun:2026agt,Canullan-Pascual:2026fyb}. Within the present limiting
comparison, the frozen result provides an upper-limit estimate of the
direct \(\Delta\)-induced buoyancy, while the strong-equilibrium result
describes the response when in-medium \(\Delta\) conversion is rapid
compared with the oscillation period.

\begin{figure}[t]
    \centering
        \includegraphics[width=0.90\linewidth]{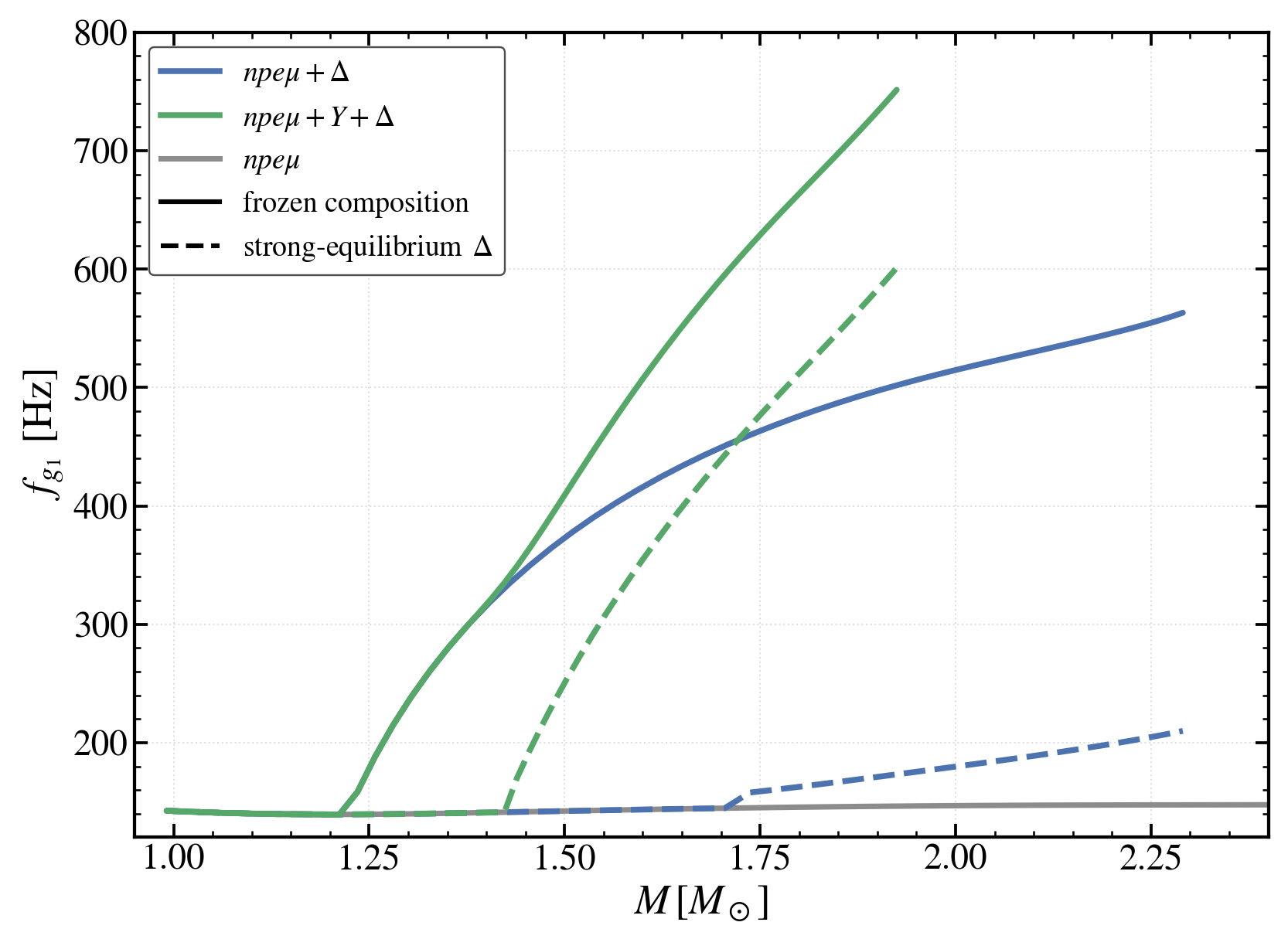}
    \caption{Full-GR $g_1$-mode frequency as a function of gravitational mass for the $npe\mu+\Delta$ and $npe\mu+Y+\Delta$ sequences. Solid curves correspond to fully frozen composition, while dashed curves are obtained by imposing strong equilibrium within the $\Delta$ quartet at fixed lepton fractions and strangeness. The nucleonic $npe\mu$ sequence is shown for reference. }
    \label{fig:strong_D}
\end{figure}

\cref{fig:strong_D} compares the full-GR
\(g_1\)-mode frequencies obtained in the fully frozen and
strong-equilibrium \(\Delta\) limits. When all particle fractions are
frozen, both \(\Delta\)-bearing sequences depart from the nucleonic
branch at \(M\simeq1.18\,M_\odot\). At the fixed mass
\(1.4\,M_\odot\), their frozen frequencies are approximately
\(316\)--\(317\) Hz. Allowing the \(\Delta\) quartet to maintain strong
equilibrium removes most of this enhancement and returns both modes to
approximately \(142\) Hz, near the nucleonic band, on the same
\(1.4\,M_\odot\) backgrounds. For \(npe\mu+\Delta\), the paired frozen
and strong-equilibrium frequencies at the EOS-validity terminal
configuration are \(563\) and \(210\) Hz, respectively, on the same
stellar background. The large enhancement in the frozen limit is
therefore produced by the retained
\(\Delta\)-fraction gradients, while the small residual offset from the
nucleonic sequence arises from the remaining lepton stratification,
consistent with \cref{tab:hypdelta_chi}. For
\(npe\mu+Y+\Delta\), a high-frequency branch develops above
\(M\simeq1.5\,M_\odot\) even after the direct \(\Delta\) contribution
is removed. At its EOS-validity terminal configuration,
\(M_{\rm term}=1.924\,M_\odot\), the paired frozen and
strong-equilibrium frequencies are \(751\) and \(601\) Hz,
respectively, on the same stellar background.
Within the present calculation, this surviving mode is driven by the
frozen \(\Lambda\) gradient rather than by the strongly equilibrated
\(\Delta\) quartet. The hyperon fractions are retained as frozen
variables here; because a corresponding hyperonic reaction-rate
calculation has not been performed, no claim is made that the
\(\Lambda\)-driven branch is protected under all thermodynamic
conditions. The direct \(\Delta\)-induced buoyancy is therefore strongly
suppressed in the fast-\(\Delta\) limit, whereas the kaon-associated mode
remains super-nucleonic even in the fast-\(K\) limit. Thus, within the
present model, the \(\Delta\) sector alone cannot sustain a mode
substantially above the nucleonic band when strong equilibrium is
maintained.

\subsection{Resonant tidal excitation in binary inspirals}
  \label{sec:tidal}

During a binary inspiral, the quadrupolar \(m=2\)
tidal field resonantly excites the \(g_1\)-mode when
\(2\Omega_{\rm orb}=\omega_{g_1}^{\rm GR}\), or equivalently when
\(f_{\rm GW}=f_{g_1}^{\rm GR}=f_{\rm GR}\). The resonance imprints a phase shift
\(\Delta\Phi_{g_1}\) on the gravitational waveform
\cite{1994ApJ...426..688R,Lai:1993di}.

\begin{figure}[t]
  \centering
  \includegraphics[width=\columnwidth]{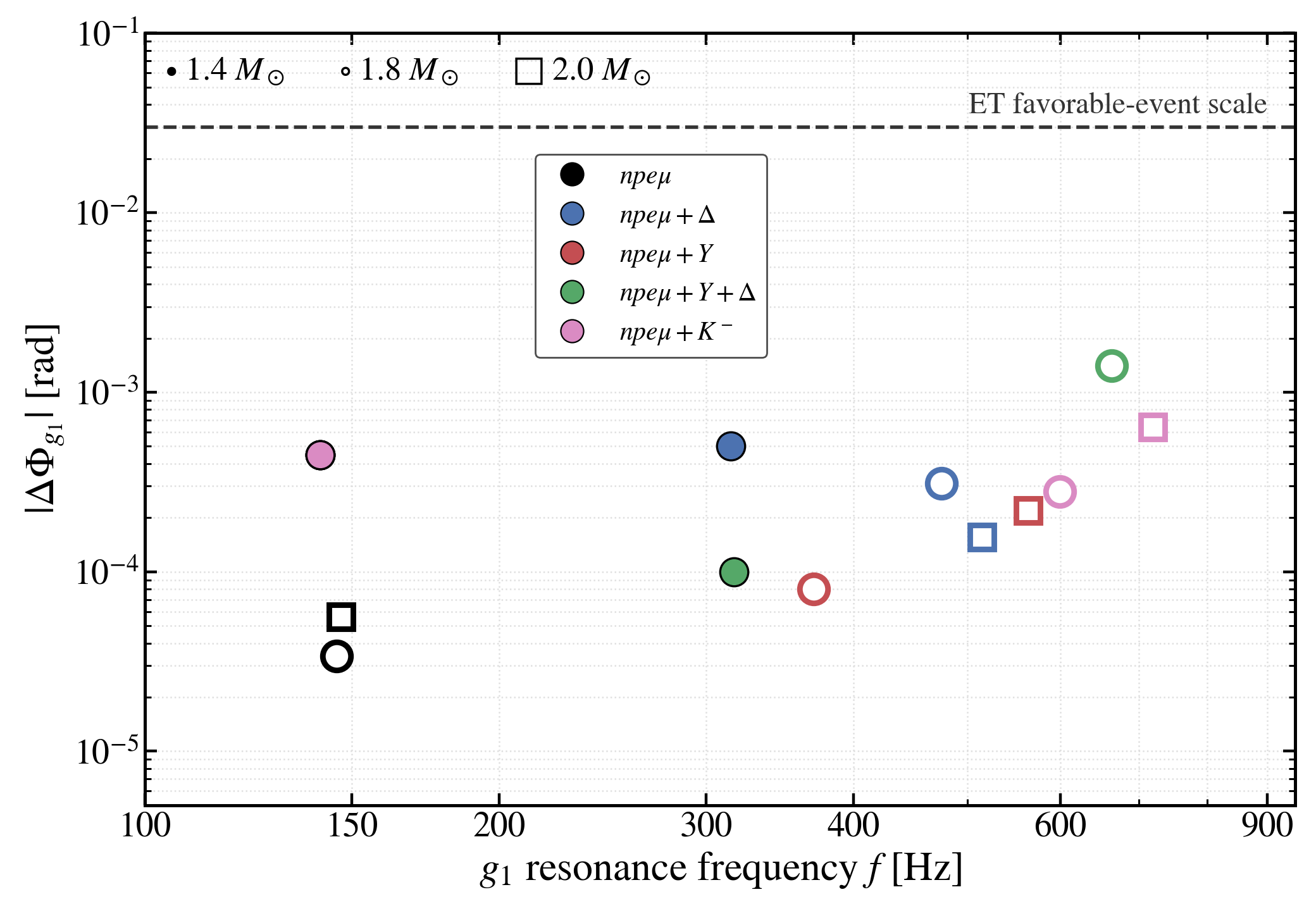}
 \caption{Resonant tidal phase shift
\(|\Delta\Phi_{g_1}|\) versus \(g_1\)-mode resonance frequency for
equal-mass binaries, shown at \(1.4\) (filled circles), \(1.8\) (open
circles), and \(2.0\,M_\odot\) (open squares). Colors denote the
composition; the antikaon sequence is shown for \(U_K=-160\) MeV.
The overlaps, phase shifts, and resonance frequencies are obtained
directly from full-GR mode solutions with frozen composition; no Cowling
overlap or empirical calibration is applied. The dashed line marks
\(\Delta\Phi\simeq0.03\) rad, the smallest phase shift found to be
identifiable for favorable Einstein Telescope events in
Ref.~\cite{Gittins:2026ntx}; it is not a universal detection threshold.
No \(2.0\,M_\odot\) point is shown for the
\(npe\mu+Y+\Delta\) sequence because the calculated sequence terminates
at the EOS-validity limit \(M_{\rm end}=1.924\,M_\odot\), before a
physical maximum is reached.}
  \label{fig:dphi}
  \end{figure}

For \(l=2\), the full-GR Lindblom--Detweiler eigenfunctions define the
real displacement and Lagrangian-pressure variables
\[
\begin{aligned}
\xi^r&=r^{l-1}e^{-\lambda/2}\operatorname{Re}W,\\
V_c&=-r^l\operatorname{Re}V,\\
\Delta P&=-r^l e^{-\nu/2}\operatorname{Re}X .
\end{aligned}
\]
The Eulerian energy-density perturbation used in the overlap is
\[
\delta\varepsilon=
\frac{\varepsilon+P}{\Gamma_1P}\Delta P
-\xi^r\frac{d\varepsilon}{dr},
\]
where \(\Gamma_1\) is the frozen-composition adiabatic index and
\(d\varepsilon/dr=(dP/dr)/c_e^2\). With the manuscript metric
convention \(g_{tt}=-e^\nu\) and \(g_{rr}=e^\lambda\), the implemented
signed dimensional overlap and fluid mode norm are
\[
\begin{aligned}
Q_{\rm GR}={}&\int_0^R
\delta\varepsilon\,r^l e^{(\lambda+\nu)/2}r^2\,dr,\\
A_2^{\rm GR}={}&\int_0^R
(\varepsilon+P)e^{-\nu}e^{(\lambda+\nu)/2}r^2
\\
&\quad\times
\left[e^\lambda(\xi^r)^2+
\frac{l(l+1)V_c^2}{r^2}\right]
\,dr .
\end{aligned}
\]
Both integrals extend from the center to the surface of the fluid
interior; no exterior contribution enters the overlap. They are
evaluated directly from the full-GR eigenfunctions at the refined
full-GR eigenfrequency. The amplitude-invariant normalized overlap and
dimensionless frequency are
\[
\hat q_{\rm GR}=\frac{|Q_{\rm GR}|}
{\sqrt{A_2^{\rm GR}MR^2}},
\qquad
\widetilde\omega_{\rm GR}=\omega_{g_1}^{\rm GR}
\sqrt{\frac{R^3}{M}} .
\]
Here \(G=c=1\), so \(M\) and \(R\) are expressed as lengths. An
arbitrary common eigenfunction normalization scales \(Q_{\rm GR}\)
linearly and \(A_2^{\rm GR}\) quadratically and therefore cancels from
\(\hat q_{\rm GR}\). In the normalization of
Refs.~\cite{Counsell:2024pua,Gittins:2026ntx}, the implemented
equal-mass phase shift is
\[
|\Delta\Phi_{g_1}|=
\frac{5\pi^2}{1024}
\left(\frac{R}{M}\right)^5
\frac{\hat q_{\rm GR}^{\,2}}{\widetilde\omega_{\rm GR}^{\,2}}.
\]
For unequal masses, the corresponding leading-order rescaling is
\(2q/(1+q)\), with \(q=m_2/m_1\); it is not applied to
\cref{fig:dphi}.

Every point in \cref{fig:dphi} is a direct full-GR result computed from
frozen-composition eigenfunctions; we did not evaluate fast-\(K\) or
strong-\(\Delta\) tidal overlaps, so the plotted phase shifts are
frozen-composition values and no reaction-limit ordering is implied.

\cref{fig:dphi} places the \(1.4\), \(1.8\),
and \(2.0\,M_\odot\) configurations of all five compositions in the
plane of resonance frequency and tidal phase shift, with the corresponding values collected
in \cref{tab:tidal}. At \(1.4\,M_\odot\) the \(npe\mu\), \(npe\mu+Y\),
and \(npe\mu+K^-\) stars lie below their exotic-particle onsets and
coincide at the leptonic mode (\(139.6\) Hz), whereas the two
\(\Delta\)-bearing stars have already crossed the \(\Delta^-\) threshold
and sit near \(315\) Hz. As the mass increases, each exotic species
drives its resonance to higher frequency, so by \(1.8\)--\(2.0\,M_\odot\)
the five compositions spread across roughly \(145\)--\(720\) Hz and
separate cleanly. The phase shifts stay small throughout,
\(|\Delta\Phi_{g_1}|\sim10^{-5}\)--\(10^{-3}\) rad, and do not track the
frequency monotonically: the largest belongs to the \(npe\mu+Y+\Delta\)
mode near \(1.8\,M_\odot\), the smallest to the low-frequency nucleonic
mode. The \(npe\mu+Y+\Delta\) sequence has no \(2.0\,M_\odot\) entry
because it terminates at its EOS-validity configuration,
\(M_{\rm end}=1.924\,M_\odot\).

\begin{table*}[t]
\centering
\caption{Resonance frequencies \(f_{g_1}\) and frozen-composition tidal
phase shifts \(|\Delta\Phi_{g_1}|\) for equal-mass binaries at
\(1.4\), \(1.8\), and \(2.0\,M_\odot\), for the five compositions of
\cref{fig:dphi} (antikaon sequence at \(U_K=-160\) MeV). At
\(1.4\,M_\odot\) the \(npe\mu\), \(npe\mu+Y\), and \(npe\mu+K^-\) stars
lie below their exotic-particle onsets and share the leptonic mode. The
\(npe\mu+Y+\Delta\) sequence terminates at \(1.924\,M_\odot\), so no
\(2.0\,M_\odot\) entry exists.}
\label{tab:tidal}
\begin{ruledtabular}
\begin{tabular}{l cc cc cc}
 & \multicolumn{2}{c}{\(1.4\,M_\odot\)}
 & \multicolumn{2}{c}{\(1.8\,M_\odot\)}
 & \multicolumn{2}{c}{\(2.0\,M_\odot\)}\\
\cline{2-3}\cline{4-5}\cline{6-7}
Composition
 & \(f_{g_1}\) & \(|\Delta\Phi_{g_1}|\)
 & \(f_{g_1}\) & \(|\Delta\Phi_{g_1}|\)
 & \(f_{g_1}\) & \(|\Delta\Phi_{g_1}|\)\\
 & (Hz) & (rad) & (Hz) & (rad) & (Hz) & (rad)\\
\hline
\(npe\mu\)
 & \(139.6\) & \(4.49\times10^{-4}\)
 & \(145.0\) & \(3.4\times10^{-5}\)
 & \(146.2\) & \(5.6\times10^{-5}\)\\
\(npe\mu+Y\)
 & \(139.6\) & \(4.49\times10^{-4}\)
 & \(370.1\) & \(8.0\times10^{-5}\)
 & \(563.9\) & \(2.20\times10^{-4}\)\\
\(npe\mu+K^-\)
 & \(139.6\) & \(4.49\times10^{-4}\)
 & \(597.0\) & \(2.80\times10^{-4}\)
 & \(716.8\) & \(6.40\times10^{-4}\)\\
\(npe\mu+\Delta\)
 & \(314.8\) & \(5.00\times10^{-4}\)
 & \(475.3\) & \(3.10\times10^{-4}\)
 & \(514.4\) & \(1.56\times10^{-4}\)\\
\(npe\mu+Y+\Delta\)
 & \(316.6\) & \(1.00\times10^{-4}\)
 & \(662.9\) & \(1.410\times10^{-3}\)
 & \multicolumn{2}{c}{--}\\
\end{tabular}
\end{ruledtabular}
\end{table*}

The largest resolved phase shift,
\(|\Delta\Phi_{g_1}|=1.410\times10^{-3}\) rad for \(npe\mu+Y+\Delta\)
near \(1.8\,M_\odot\), is about a factor of \(21\) below the
\(0.03\)-rad favorable-event scale quoted for the Einstein Telescope
\cite{Gittins:2026ntx}. That scale is not a universal detection
threshold, and the comparison does not establish detectability for the
present modes; within the circular, equal-mass, nonrotating,
frozen-composition assumptions used here it simply quantifies their
small direct full-GR tidal response. The antikaon case also illustrates
that gravitational-wave damping and tidal excitation probe different
properties of a mode: its damping time drops sharply after condensation
[\cref{fig:kaon_gmode}], while its small tidal phase shift reflects weak
coupling to the external tidal field.

Two implications follow. First, in the adopted
normalization, the smooth composition modes considered here have small
tidal overlaps: the full resolved set spans
\(4.32\times10^{-5}\leq\hat q_{\rm GR}\leq1.31\times10^{-3}\), while
the subset classified as appreciable spans
\(1.02\times10^{-4}\leq\hat q_{\rm GR}\leq1.31\times10^{-3}\).
The phase shifts satisfy
\(|\Delta\Phi_{g_1}|\leq1.410\times10^{-3}\) rad. By contrast, published
calculations of discontinuity modes associated with first-order phase
transitions obtain phase shifts as large as
\(|\Delta\Phi|\sim0.1\)--\(1\) rad at comparable frequencies
\cite{Miao:2023jqe,Pereira:2025xsi}. A large measured phase shift in
the \(300\)--\(700\) Hz band would therefore favor a sharp internal
interface over the smooth composition stratification studied here and
could help distinguish between crossover and first-order-transition
scenarios. It would not, however, constitute a unique diagnostic,
because rotation, orbital eccentricity, and finite-rate composition
changes can also modify the resonant response. Second, recent
calculations show that moderate orbital eccentricities,
\(e_{\rm 10\,Hz}\sim0.2\)--\(0.4\), can enhance the detectability of
\(g\)-mode dynamical tides by more than an order of magnitude
\cite{Takatsy:2026doc}. The circular-binary phase shifts reported here
therefore provide a baseline for future calculations that include
rotation, eccentricity, and reaction-rate effects, rather than direct
predictions for eccentric systems.

\section{Conclusions and outlook}
\label{sec:conclusions}

We have presented, to our knowledge, the first calculation in full general
relativity of the continuous-composition core \(g_1\)-mode frequency and
gravitational-wave damping time in neutron stars containing a \(K^-\)
condensate, together with a unified comparison with hyperonic and
\(\Delta\)-admixed matter. Antikaon condensation creates a localized buoyancy
layer and a distinct high-frequency branch. Fast-\(K\) equilibration retains
approximately \(36\%\)--\(44\%\) of
the peak local buoyancy and \(65.7\%\)--\(73.4\%\) of the frozen
terminal-configuration frequencies, while lengthening the gravitational-wave
damping times by factors of \(14.4\)--\(31.8\); the mode nevertheless remains
above the nucleonic band. The species-resolved Ledoux decomposition and
frequency sensitivities identify the kaon channel as its dominant driver. By
contrast, strong equilibrium within the \(\Delta\) quartet removes most of the
direct \(\Delta\)-induced enhancement and returns the \(N\Delta\) mode toward
the nucleonic band, whereas the high-frequency \(NY\Delta\) branch survives
because the \(\Lambda\) gradient is kept frozen. Full-GR eigenfunction tracking
confirms that these changes occur along the same continuous \(g_1\) branch.

The direct full-GR tidal calculation with frozen-composition eigenfunctions
gives \(|\Delta\Phi_{g_1}|\leq1.410\times10^{-3}\) rad, about a factor of 21
below the \(0.03\)-rad favorable-event scale quoted for the Einstein Telescope.
For the circular, equal-mass, nonrotating models considered here, this weak
response contrasts with the larger phase shifts predicted for some
discontinuity modes, but it is not a unique diagnostic of a first-order
transition. Representative DD-ME2 calculations reproduce the same
reaction-channel hierarchy despite quantitative EOS dependence; agreement
between two functionals supports robustness but does not establish
universality. The reaction prescriptions are limiting cases, the reported
damping times include gravitational radiation only, the
hyperon fractions remain frozen, and no fast-\(K\) or strong-\(\Delta\) tidal
overlap has been calculated. The immediate next step is a finite-rate kinetic
treatment with chemical damping and reaction-consistent tidal overlaps,
followed by extensions to rotating and eccentric binaries.

\begin{acknowledgments}
P.~Thakur is supported by the National Research Foundation of Korea (NRF) grant funded by the Korea government (MSIT) (No.~RS-2024-00457037). This work was supported (in part) by the Yonsei University Research Fund(Yonsei University Frontier Fellowship for Postdoctoral Researchers) of 2025. I.~A.~R. gratefully acknowledges support from the Deutsche Forschungsgemeinschaft (DFG, German Research Foundation) – Project Number 579861443 and also in part by the Alexander von Humboldt Foundation through a Humboldt Research Fellowship.
\end{acknowledgments}

\section*{Data Availability}
The data that support the findings of this study are available from the corresponding author upon reasonable request.

\appendix

\section{DD-ME2 test of EOS dependence}
\label{app:ddme2}

To test whether the reaction-limit conclusions depend on the nucleonic
EOS, we repeat representative calculations with the density-dependent
DD-ME2 functional \cite{Lalazissis:2005de}, which is widely used for
unified neutron-star structure and for hyperonic and
\(\Delta\)-admixed matter
\cite{Xia:2022dvw,Thapa:2021ifv,Tran:2022dva,Rather:2024mtd}. We retain the exotic-particle prescriptions adopted in the main
calculation while recalibrating the scalar couplings fixed by
saturation-density potentials. For the hyperons, the scalar couplings
are recalibrated to
\(U_\Lambda=-28\) MeV, \(U_\Sigma=+30\) MeV, and
\(U_\Xi=-14\) MeV, while the SU(6) vector-coupling relations, including
the \(\phi\) meson, are retained. For the antikaon calculation, we use
\(U_K=-120\) MeV,
\(g_{\omega K}=g_{\omega N}/3\), and
\(g_{\rho K}=g_{\rho N}\), with \(g_{\sigma K}\) recalibrated using the
DD-ME2 saturation fields. For the \(\Delta\) baryons, the fixed ratios
\(x_{\sigma\Delta}=1.09\), \(x_{\omega\Delta}=1.05\), and
\(x_{\rho\Delta}=2.5\) yield
\(U_\Delta(n_0)=-99.25\) MeV in DD-ME2; this value is a prediction of
the adopted ratios rather than a calibration input.

For \(i=\sigma,\omega,\rho\), the exotic-particle couplings inherit the
density dependence of the corresponding nucleonic DD-ME2 coupling,
\[
g_{iX}(n_B)=x_{iX}g_{iN}(n_B),
\]
where \(n_B\) is the total baryon density and excludes the kaon
condensate. For the hidden-strangeness interaction, for which there is
no nucleonic \(\phi\) coupling, we use
\(g_{\phi Y}(n_B)=x_{\phi Y}g_{\omega N}(n_B)\). The native DD-ME2
isovector convention, \(\tau_3=2I_3\), is used throughout, and the
rearrangement self-energy generated by the density-dependent couplings
is included consistently in the chemical potentials and pressure.

The implementation reproduces the DD-ME2 saturation properties
\(n_0=0.1519\,\mathrm{fm}^{-3}\),
\(E/A=-16.14\) MeV, \(K=250.9\) MeV,
\(J=32.30\) MeV, and \(L=51.27\) MeV. It satisfies thermodynamic
consistency, beta equilibrium, charge neutrality, and baryon-number
conservation. The representative sequences are continuous, causal,
mechanically stable, and convex, and reach physical maximum masses
above \(2\,M_\odot\). We calculate \(c_s^2-c_e^2\) and the full-GR
\(g_1\)-mode frequency using the same procedure as in the main text,
holding each TOV background fixed when comparing the frozen and
reaction-equilibrated limits. The paired limits coincide below the
relevant particle threshold, and radial-node tracking confirms that
the post-onset solutions remain on continuous \(g_1\) branches.

\begin{figure*}[t]
    \centering
    \includegraphics[width=\textwidth]
    {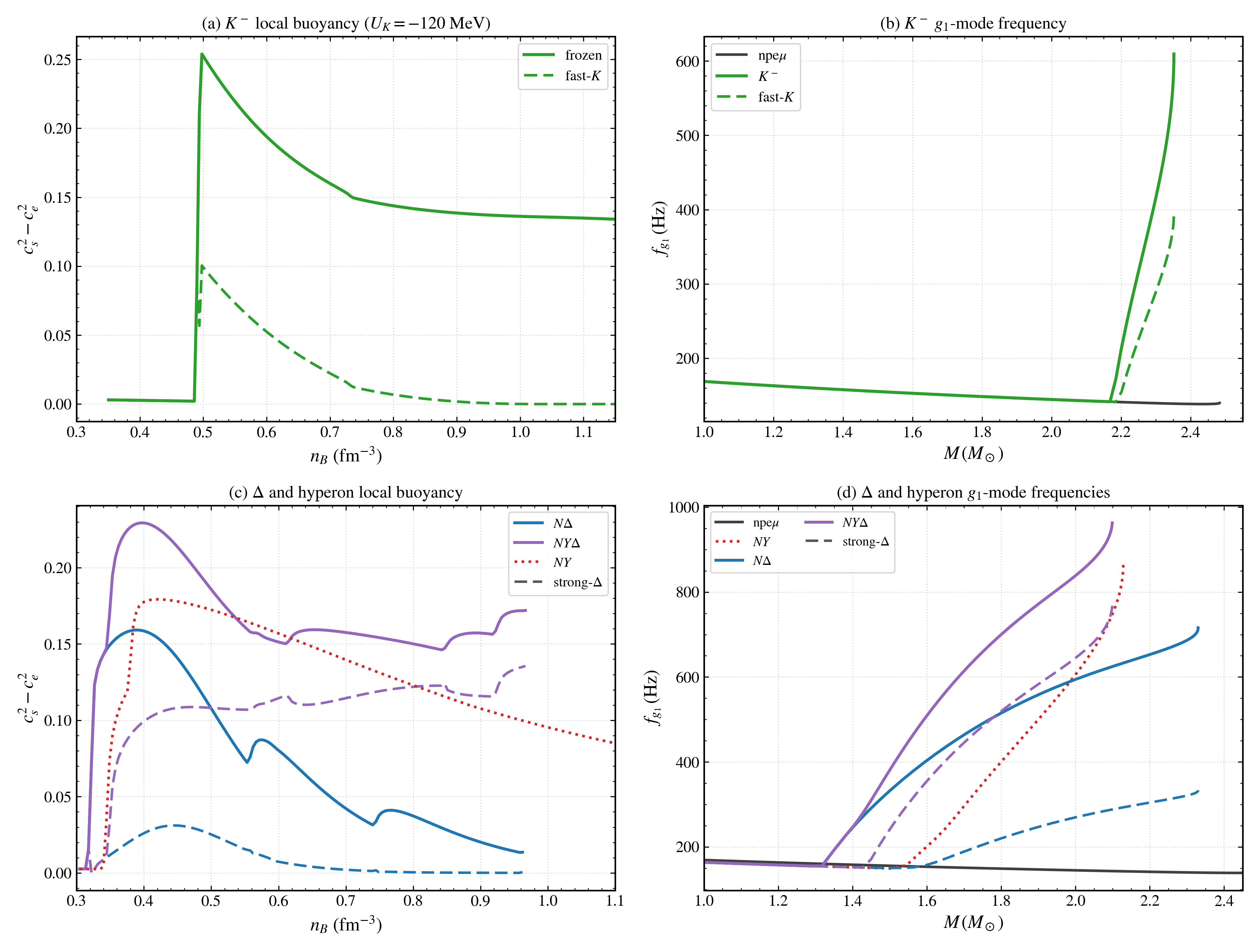}
    \caption{DD-ME2 test of the reaction-constrained composition
    buoyancy. Panel (a) shows the local buoyancy measure
    \(c_s^2-c_e^2\) for \(K^-\)-condensed matter with
    \(U_K=-120\) MeV, while panel (b) shows the corresponding full-GR
    \(g_1\)-mode frequencies. The solid and dashed green curves denote
    the frozen and fast-\(K\) limits, respectively, and the black curve
    is the nucleonic reference. Panel (c) compares the local buoyancy
    of the \(N\Delta\) and \(NY\Delta\) compositions in the frozen and
    strong-\(\Delta\) limits, together with the frozen \(NY\) reference.
    Panel (d) shows their corresponding full-GR \(g_1\)-mode
    frequencies. Blue and purple identify the \(N\Delta\) and
    \(NY\Delta\) sequences, respectively; solid curves denote frozen
    composition and dashed curves denote strong \(\Delta\) equilibrium.
    The dotted red curve is the frozen \(NY\) reference, and the black
    curve is the nucleonic sequence. For each reaction-limit
    comparison, the equilibrium EOS and TOV background are identical;
    only the perturbative composition constraint is changed.}
    \label{fig:ddme2_robustness}
\end{figure*}

\Cref{fig:ddme2_robustness,tab:ddme2_comparison} show that the three
reaction-limit mechanisms found with BigApple persist with DD-ME2.
For the antikaon case, the fast-\(K\) frequency retains \(64.9\%\) of
the frozen value with DD-ME2, compared with \(66.6\%\) for BigApple,
and remains well above the corresponding nucleonic band. The retained
peak local buoyancy is \(40\%\) for DD-ME2 and \(39.6\%\) for
BigApple.

For \(N\Delta\) matter, strong \(\Delta\) equilibration returns the
mode close to the nucleonic band near \(1.4\,M_\odot\), while at
\(1.9\,M_\odot\) it retains \(44.2\%\) of the frozen frequency with
DD-ME2, compared with \(34.4\%\) for BigApple. The corresponding
retained peak local buoyancies are \(15\%\) and \(16.3\%\),
respectively. For \(NY\Delta\) matter, the high-frequency branch
survives strong \(\Delta\) equilibration, retaining \(75.5\%\) of the
frozen frequency with DD-ME2 and \(79.5\%\) with BigApple. The retained
peak local buoyancies are \(43\%\) and \(46.1\%\), respectively. The
surviving response is carried primarily by the frozen \(\Lambda\)
gradient rather than by the equilibrated \(\Delta\) quartet, as also
confirmed by the species-sensitivity diagnostic. This attribution
applies under the frozen-hyperon assumption used here; finite-rate
hyperonic equilibration has not been calculated.

The absolute differences between the two EOSs in the
frequency-retention ratios are \(0.017\), \(0.098\), and \(0.040\) for
the \(K^-\), \(N\Delta\), and \(NY\Delta\) cases, respectively. These
quantitative differences reflect the density dependence of the
couplings and the ordering of the particle thresholds. Nevertheless,
the same qualitative reaction-channel hierarchy is obtained with both
EOSs: kaon buoyancy is partially retained in the fast-\(K\) limit,
direct \(\Delta\)-induced buoyancy is strongly suppressed by
strong-equilibrium conversion, and the \(NY\Delta\) branch survives
through the frozen hyperon gradient. A broader EOS survey would be
required to establish complete model independence.

\begin{table*}[t]
\centering
\caption{Comparison of the reaction-constrained full-GR \(g_1\)-mode
results obtained with BigApple and DD-ME2. Here
\(f_{g_1}^{\rm react}\) denotes the fast-\(K\) frequency for the
\(K^-\) rows and the strong-\(\Delta\) frequency for the
\(N\Delta\) and \(NY\Delta\) rows. The kaon results are evaluated at
the maximum-mass configuration of each EOS, the baryonic results at
\(1.9\,M_\odot\). The local retention
\(\mathcal{R}_{\rm loc}
=100(c_s^2-c_e^2)_{\rm react}/(c_s^2-c_e^2)_{\rm frozen}\) is evaluated
at the density where the frozen profile peaks; it is not a global
mode-buoyancy fraction.}
\label{tab:ddme2_comparison}
\begingroup
\small
\setlength{\tabcolsep}{5.5pt}
\renewcommand{\arraystretch}{1.25}
\begin{tabular}{l l c c c c c}
\hline\hline
Composition & EOS &
\(M_{\rm cmp}\) &
\(f_{g_1}^{\rm frozen}\) &
\(f_{g_1}^{\rm react}\) &
\(f_{g_1}^{\rm react}/f_{g_1}^{\rm frozen}\) &
\(\mathcal{R}_{\rm loc}\) \\
 & & \((M_\odot)\) & (Hz) & (Hz) & & (\%) \\
\hline
\(K^-\)       & BigApple & 2.413 & 651.7 & 433.7 & 0.666 & 39.6 \\
\(K^-\)       & DD-ME2   & 2.352 & 609.3 & 395.6 & 0.649 & 40 \\
\hline
\(N\Delta\)   & BigApple & 1.900 & 497.4 & 171.2 & 0.344 & 16.3 \\
\(N\Delta\)   & DD-ME2   & 1.900 & 558.2 & 246.9 & 0.442 & 15 \\
\hline
\(NY\Delta\)  & BigApple & 1.900 & 733.8 & 583.3 & 0.795 & 46.1 \\
\(NY\Delta\)  & DD-ME2   & 1.900 & 771.8 & 583.0 & 0.755 & 43 \\
\hline\hline
\end{tabular}
\endgroup
\end{table*}

\clearpage
\bibliographystyle{apsrev4-2}
\bibliography{reference}

\end{document}